\documentclass[aps,prl,twocolumn,preprintnumbers,amsmath,amssymb,superscriptaddress,10pt]{revtex4-2}
\usepackage{amsmath,graphicx}
\usepackage[utf8]{inputenc}
\usepackage[T1]{fontenc}
\usepackage{xcolor}
\usepackage{textcomp}
\usepackage{bm}

\usepackage{slashed}
\usepackage{subfigure}
\usepackage{physics}
\usepackage{amsmath}
\usepackage{tikz}
\usepackage{mathdots}
\usepackage{yhmath}
\usepackage{cancel}
\usepackage{color}
\usepackage{array}
\usepackage{multirow}
\usepackage{amssymb}
\usepackage{gensymb}
\usepackage{tabularx}
\usepackage{extarrows}
\usepackage{booktabs}
\usetikzlibrary{fadings}
\usetikzlibrary{patterns}
\usetikzlibrary{shadows.blur}
\usetikzlibrary{shapes}
\usepackage{feynmp-auto}

\usepackage{color}
\definecolor{LinkColor}{rgb}{0.235,0.235,0.588}

\usepackage{hyperref}
\hypersetup{
	pdfauthor={good guys},
	pdftitle={good title},
	colorlinks=true,
	citecolor=LinkColor,
	linkcolor=LinkColor,
	urlcolor=LinkColor,
}

\usepackage{listings}
\definecolor{lightgray}{gray}{1}

\usepackage{theorem}

\usepackage{tikz}

\newcommand{\nc}{\newcommand}
\nc{\braoprket}[3]{\langle#1|#2|#3\rangle}
\nc{\opn}[1]{\operatorname{#1}}
\nc{\avg}[1]{\langle#1\rangle}
\nc{\ketbrasame}[1]{|#1\rangle\!\langle#1|}
\nc{\swap}{\opn{SWAP}}
\nc{\E}{\mathbb{E}}
\nc{\Var}{\opn{Var}}
\nc{\dg}{\dagger}
\newcommand{\K}{\mathbf{K}}

\usepackage[normalem]{ulem}

\begin{document}

\title{Boundary Criticality in (2+1)-dimensional U(1) Dirac Quantum Spin Liquid}

\author{Huan Jiang}
\altaffiliation{The first two authors contributed equally.}
\affiliation{Department of Physics and Engineering Physics, Tulane University, New Orleans, Louisiana, 70118, USA}

\author{Zhiming Pan}
\altaffiliation{The first two authors contributed equally.}
\affiliation{Department of Physics, Xiamen University, Xiamen 361005, China}

\author{Xue-Jia Yu}
\email{xuejiayu@eitech.edu.cn}
\affiliation{Eastern Institute of Technology, Ningbo 315200, China}

\author{Shao-Kai Jian}
\email{sjian@tulane.edu}
\affiliation{Department of Physics and Engineering Physics, Tulane University, New Orleans, Louisiana, 70118, USA}

\begin{abstract}
An emergent gauge field can control boundary critical behavior without changing the bulk quantum spin liquid. 
We investigate the boundary criticality in $(2+1)$-dimensional $U(1)$ Dirac spin liquid, whose low-energy physics is described by massless quantum electrodynamics. 
Using a perturbative renormalization-group analysis in a half-space, we show that Neumann and Dirichlet boundary conditions for the emergent gauge field lead to distinct boundary universality classes. 
We determine the boundary scaling dimensions of the fundamental fields and gauge-invariant operators that provide observable signatures. 
We further propose a fermion-gauge lattice model with tunable boundary interactions as a microscopic setting for realizing and probing these boundary universality classes.
\end{abstract}

\maketitle

\emph{Introduction.}---A bulk critical point can admit inequivalent conformal boundary conditions and hence distinct boundary universality classes, whose operator spectra and scaling dimensions are not fixed by the bulk theory alone~\cite{andrei2020boundary}.
This structure is well established in bosonic $O(N)$ models, which support ordinary, special, and extraordinary surface transitions~\cite{diehl1981field,diehl1983multicritical,domb1986phase,diehl1996the,cardy1996scaling,liendo2012the,giombi2020cft,Giombi2020boundaryInteractions,metlitski2020boundary,sun2025boundary,sun2026analytic}, and has recently been extended to interacting fermionic boundary conformal field theories (BCFTs), including Gross--Neveu and Gross--Neveu--Yukawa models~\cite{Giombi2021cnr,Herzog2022bcb,jian2025bdygny,
Fedorenko2026bgny,Diatlyk2026higherOrderGNY}.
In correlated topological insulators and superconductors, boundary critical behavior is further constrained by bulk symmetry and topology~\cite{grover2014emergent,li2017edge,Ge2025ti,Ge2025tsc,shen2025boundary}.
Related work on gapless topological and symmetry-enriched critical phases has established protected edge structure and anomaly constraints~\cite{yu2026topological,Scaffidi2017PRX,Verresen2018PRL,Verresen2021PRX,Thorngren2021PRB,Duque2021PRB}, and has developed concrete boundary, entanglement, and multicritical diagnostics across spin chains, circuits, and synthetic platforms~\cite{Yu2022PRL,Yu2024PRL,Prembabu2024PRB,Yu2024PRB,Zhang2024PRA,Li2025SciPost,Huang2025SciPost,Deng2026PRL,Zhong2025PRB,Yang2026PRBL,guo2026lihaldane,prembabu2025multi,prembabu2025non,Yu2026PRB,Tan2026CP,xu2026frame,yang2026topologicaltricriticalisinguniversality,chou2026topologicallyenforcedlifshitzmulticriticality,chou2026ptsymmetryenrichednonunitarycriticality,chen2026criticaltopologicalphotonicssynthetic}.



Emergent gauge fields provide a further source of boundary dynamics. 
A natural setting for this problem is the $U(1)$ Dirac spin liquid, whose massless Dirac spinons couple to an emergent $U(1)$ gauge field and are described at low energies by (2+1) dimensional quantum electrodynamics (QED$_3$)~\cite{savary2016quantum,wen2002aQSL,DiPietro2015qed,Giombi2016currentsQED,DiPietro2015qed,DiPietro2017scalingQED3}. 
This state has been studied in square-lattice $\pi$-flux constructions and proposed for kagome- and triangular-lattice antiferromagnets~\cite{Hermele2004stability,Ran2007kagomeDSL,Hermele2008kagomeASL,He2017kagomeDSL,Hu2019triangularDSL}. 
Related Abelian gauge theories also arise in descriptions of doped Mott insulators, deconfined quantum criticality, and quantum spin ice~\cite{lee2006doping,senthil2004deconfined,wang2017deconfined,gingras2014quantum,wan2026quantum}. 
Note that gauge fields coupled to lower-dimensional matter have been studied in reduced and mixed-dimensional QED~\cite{Teber2012reducedQED,Hsiao2017mixedQED,FraserTaliente2025nonlocalSchwinger}, in graphene-inspired BCFTs~\cite{HerzogHuang2017boundaryCentral,Herzog2018grapheneBCFT}, and through conformal bootstrap analyses~\cite{DiPietro2019bdygauge,BartlettTisdall2024boundaryQED}.
Nevertheless, how gauge-field boundary conditions select distinct boundary universality classes of a $U(1)$ Dirac spin liquid, and how these conditions arise microscopically, remain largely unexplored.

In this letter, we study boundary correlation behavior in $U(1)$ Dirac spin liquid, described in low energies by QED$_3$, using an expansion near four spacetime dimensions.
Here, we formulate a one-loop renormalization group (RG) for QED on a flat half-space.
The Dirac fermions exhibit a reflecting boundary condition, while the gauge field maintain Neumann or Dirichlet boundary conditions.
With propagators compatible to the presence of suitable boundary conditions, we analyze the renormalization of boundary fields and fermion bilinears, characterizing the boundary criticality.
Finally, we discuss a lattice fermion-gauge model and the observables that probe its boundary theory, which is accessible to quantum Monte Carlo simulations~\cite{Xu2018qmc,Feng2026scalable} and potentially cold atom platform~\cite{Celi2019lqy}.

\begin{figure}[t]
\centering
\includegraphics[width=\columnwidth]{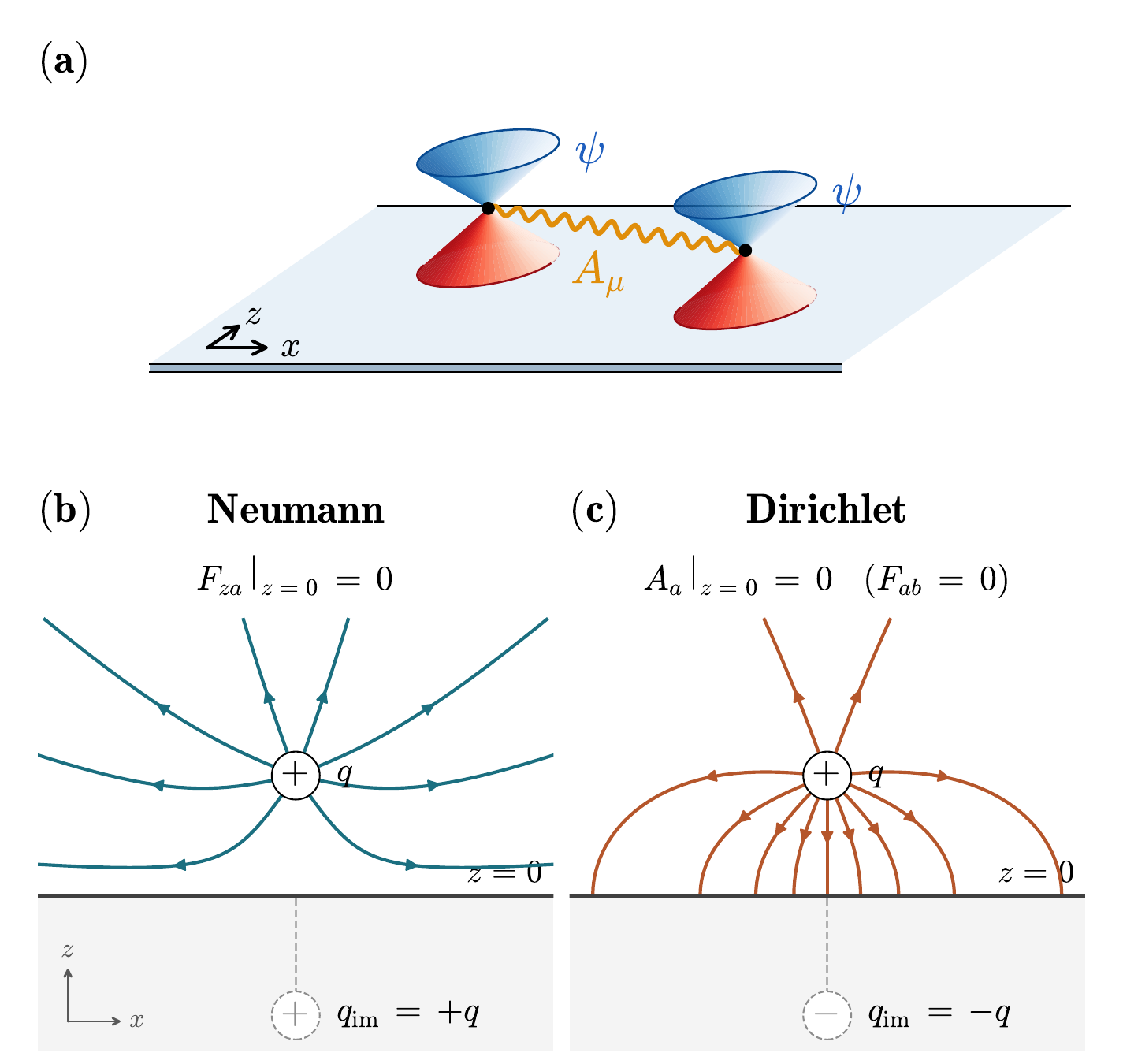}
\caption{(a) Fractionalized Dirac spinons $\psi$ in a (2+1)D quantum spin liquid coupled to a gauge field $A_\mu$, with the system subject to open boundary conditions in the $z$ direction. 
(b,c) Electrostatic image-charge picture of the Neumann (b) and Dirichlet (c) boundary conditions, corresponding to same-sign and opposite-sign image charges, respectively.}
\label{fig:ImageCharge}
\end{figure}

\emph{QED on a half-plane.}---We consider massless QED in a $d$-dimensional Euclidean spacetime within a spatial boundary. 
As illustrated in Fig.~\ref{fig:ImageCharge}, the theory consists of $N$ flavors four-component Dirac fermions $\psi_j(x)$, with ($j=1,\cdots,N$), coupled to an Abelian $U(1)$ gauge field $A_\mu(x)$.
Both fields occupy the half-plane $\mathcal{M}^+=\{x_{\mu}|z>0\}$, where we set the coordinate perpendicular to the boundary as $z=x_{d-1}$.
The minimal coupling action is $S=S_{\psi}+S_A$, with
\begin{equation}
    S_{\psi}=\int_{\mathcal{M}^+} {\rm d}^{d}x\sum_{j}^{N}\bar\psi_j\gamma^\mu\left(\partial_\mu-ieA_{\mu}\right)\psi_j,
\label{eq:fermion_action}
\end{equation}
where $e$ is the gauge coupling, $\bar\psi_j=\psi_j^\dagger\gamma^0$, 
and Dirac matrices $\gamma^\mu$ satisfying $\{\gamma^\mu,\gamma^\nu\}=2\delta^{\mu\nu}$. 
The dynamics of the gauge field is governed by the standard Maxwell action,
\begin{align}
S_{A}=\int_{\mathcal{M}^+}{\rm d}^{d}x\frac14F_{\mu\nu}F_{\mu\nu},
\end{align}
with gauge field strength tensor $F_{\mu\nu}=\partial_\mu A_\nu-\partial_\nu A_\mu$. 
For the fermions, we impose the reflecting boundary condition $\left.(1+\gamma^z)\psi_j\right|_{z=0}=0$, which enforces a vanishing normal $U(1)$ current, $\left.j_z\right|_{z=0}=\left.\sum_i\bar{\psi}_i\gamma^z\psi_i\right|_{z=0}=0$.

We employ a perturbative RG analysis within $4-\varepsilon$ dimensional regularization. 
The bare fields and couplings are related to their renormalized quantities through $\psi=\sqrt{Z_2}\psi_R$, $ A_\mu=\sqrt{Z_3}A_{\mu, R}$ and $ e=\mu^{\varepsilon/2}Z_1Z_{2}^{-1}Z_{3}^{-1/2}e_{R}$.
Here, $\mu$ is the renormalization scale. 
Gauge invariance enforces the Ward-Takahashi identity and ensures equal fermion and vertex renormalization
constants, $Z_1=Z_2$. 

Deep in the bulk regime, the boundary effect can be omitted and the ultraviolet behavior is controlled by conventional translational-variant QED. 
In Feynman gauge, the one-loop factors are $Z_2=1-{e^2}/{8\pi^2\varepsilon}$ and $Z_3=1-{Ne^2}/{6\pi^2\varepsilon}$, and the infrared fixed point is $e^{*2}=6\pi^2\varepsilon/N$~\cite{DiPietro2015qed,Giombi2016conformalQED}. 
The bulk calculation is summarized in the Supplemental Material.

By contrast, correlation functions with external points on the boundary can contain additional ultraviolet poles. 
Boundary field renormalization accounts for these surface singularities, leading to boundary behavior distinct from that in the bulk. 
This distinction is central to our study of the edge of a $U(1)$ Dirac spin liquid, where the boundary conditions constrain the allowed spinon and gauge fluctuations and thereby determine the boundary critical behavior.

\begin{table*}[t]
    \centering
    {
    \setlength{\tabcolsep}{10pt}
    \begin{tabular}{l c c c}
    \hline\hline
    System
    & $\Delta_3^{\partial}$
    & $\Delta_2^{\partial}$
    & $\eta_{\mathcal{O}}^{\partial}$
    \\
    \hline
    $(2+1)$D Dirac QSL (BC-I)
    & $1+\frac{\varepsilon}{4}$
    & $\frac32+\frac{15-4N}{8N}\varepsilon$
    & $-\frac{3}{4N}\varepsilon$
    \\[1pt]
    $(2+1)$D Dirac QSL (BC-II)
    & $1+\frac{\varepsilon}{4}$
    & $\frac32+\frac{9-4N}{8N}\varepsilon$
    & $\frac{3}{4N}\varepsilon$
    \\[1pt]
    \hline\hline
    \end{tabular}}
    \caption{Boundary scaling dimensions in Dirac quantum spin liquid (QSL). In all cases, the Dirac fermions obey the conformal boundary condition $(1+\gamma^z)\psi|_{z=0}=0$. BC-I (BC-II) denotes Neumann (Dirichlet) boundary condition for the gauge field.  
    The $\Delta_{2}^{\partial}$ and $\Delta_3^{\partial}$ are the scaling dimensions of the boundary fermion and gauge field, respectively. The $\eta_{\mathcal{O}}^{\partial}$ is the anomalous dimension of the boundary fermion composite operator $\hat{\mathcal{O}}=\bar\psi_j\gamma^5\psi_j|_{z=0}$.}
    \label{tab:scaling-dim}
    \vspace{2pt}
\end{table*}

\emph{Boundary conditions.}---
The suitable  boundary conditions for the gauge field follow from the surface term in the variation of the Maxwell action $S_A$ with respect to infinitesimal change of gauge field:
\begin{equation}
\begin{aligned}
\delta S_{A}
=-\int_{\mathcal{M}^+}{\rm d}^dx\partial_\mu F_{\mu\nu}\delta A_\nu
+\int_{\partial\mathcal{M}^+}{\rm d}^{d-1}xn_\mu F_{\mu\nu}\delta A_\nu.
\end{aligned}
\end{equation}
The first bulk term contributes to the gauge-field equation of motion, $\partial_\mu F_{\mu\nu}=0$;
variation of $S_{\psi}$ further supplies the fermion current.
The second surface term pairs the tangential gauge potential $A_a$ with the normal field strength $n_{\mu} F^{\mu a}$,
where $a$ labels the spacetime directions parallel to the boundary.

\begin{figure}[t!]
    \centering
    
    \subfigure[]{
    	\includegraphics[width=0.4\columnwidth]{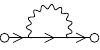}
    	\label{fig:fermion_self}
    }
    \hspace{0.05\textwidth}
    \subfigure[]{
    	\includegraphics[width=0.4\columnwidth]{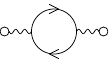}
    	\label{fig:boson_self}
    }
    \hspace{0.05\textwidth}
    \subfigure[]{
    	\includegraphics[width=0.28\columnwidth]{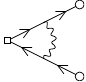}
    	\label{fig:spinor}
    }
    \caption{One-loop diagrams for the boundary renormalization of (a) the fermion self-energy, (b) the gauge field polarization, and (c) the fermion composite operator. All external legs terminate on the boundary, as indicated by $\circ$. The solid line with arrow denotes the fermion propagator, while the waving line represents the gauge field porpagator. The $\square$ represents a composite operator insertion on the boundary.} 
    \label{fig:FeynmanDiagrams}
\end{figure}

We consider homogeneous Neumann and Dirichlet boundary conditions~\cite{DiPietro2019bdygauge}.
Neumann conditions impose $F_{za}|_{z=0}=0$, and Dirichlet conditions fix $A_a|_{z=0}=0$, giving $F_{ab}|_{z=0}=0$.
In $2+1$ dimensions, these conditions become $E_z=B=0$ and $E_x=0$, respectively. 
The electric field lies along a Neumann boundary and meets a Dirichlet boundary normally.
Both conditions give zero normal Poynting flux.
Their electrostatic responses correspond to same-sign and opposite-sign image charges, respectively, as illustrated in Fig.~\ref{fig:ImageCharge} (b,c).
Note that both choices are compatible with the $U(1)$ gauge structure. 
Neumann boundary conditions allow local gauge transformations at the boundary. 
For Dirichlet boundary conditions, preservation of $A_a|_{z=0}=0$ requires $\alpha|_{z=0}$ to be constant, where $\alpha$ is the gauge-transformation parameter.
Interestingly, this constant transformation corresponds to a global boundary $U(1)$ symmetry~\cite{thorngren2023higgs,verresen2024higgs,chung2024higgs}.  
A physical realization of this boundary condition is obtained by introducing a charged boundary phase field $\varphi$, which transforms as $\varphi\to\varphi+e\alpha$. 
Its stiffness penalizes the gauge-invariant combination $(\partial_a\varphi-eA_a)^2$ and, at large stiffness, locks $e A_a$ to $\partial_a\varphi$. 
Locally choosing the unitary gauge $\varphi=0$ then gives the desired Dirichlet boundary condition, $A_a|_{z=0}=0$. 
In the lattice model below, Eq.~\eqref{eq:bdy_hamiltonian} provides the corresponding compact realization. 

\emph{Boundary universality class.}---Within Feynman gauge, the gauge field propagator can be constructed using the method of images~\cite{nishioka2022method}. 
In the mixed representation, the gauge field propagator takes the form
\begin{align}
D_{\mu\nu}(k,z,z')&=D_{\mu\nu}^{(b)}(k,z,z')\pm R_{\mu\rho}D_{\rho\nu}^{(b)}(k,\bar{z},z'),\nonumber\\
D_{\mu\nu}^{(b)}(k,z,z')&=\frac{\delta_{\mu\nu}}{2q}e^{-q|z-z'|},
\end{align}
where $q^2=\sum_{\mu\neq z}k_\mu^2$ is the squared tangential momentum and $\bar{z}=-z$ is the image coordinate.
The direct term $D_{\mu\nu}^{(b)}$ is the translational-invariant bulk propagator.
The reflection matrix $R_{\mu\nu}=\delta_{\mu\nu}-2\delta_{z\mu}\delta_{z\nu}$ reverses the normal vector component.
The upper (plus) and lower (minus) signs give Neumann and Dirichlet boundary conditions, respectively.

The fermion propagator follows from the same image construction~\cite{jian2025bdygny}. 
The boundary condition $(1+\gamma^z)\psi_j|_{z=0}=0$ gives
\begin{align}
G(k,z,z')&=G^{(b)}(k,z,z')-\gamma^zG^{(b)}(k,\bar z,z'),\nonumber\\
G^{(b)}(k,z,z')&=\left(\frac{i\slashed{k}}{2q}-\frac{\gamma^z}{2}\text{sgn}(z-z')\right)e^{-q|z-z'|},
\end{align}
where $\slashed{k}=\sum_{\mu\neq z}k_\mu\gamma^\mu$.
Here, $G^{(b)}(k,z,z')$ denotes the bulk propagator in the abscent of the boundary. 
The matrix $-\gamma^z$ in the second term acts on the reflected fermion line and enforces the boundary projection.

We denote the boundary fermion by $\hat\psi_j=\psi_j|_{z=0}$ and the fluctuating boundary gauge components by $\hat A_\mu=A_\mu|_{z=0}$.
For Neumann boundary conditions these are the tangential components; for Dirichlet boundary conditions the nonzero boundary potential in this gauge is $A_z$.
Their renormalization factors are defined by $\hat\psi=\sqrt{Z_2Z_2^\partial}\hat\psi_R$ and $\hat A_\mu=\sqrt{Z_3Z_3^\partial}\hat A_{\mu,R}$, with the gauge-field relation applied to the component under consideration.
Here, $Z_2^\partial$ and $Z_3^\partial$ absorb the additional ultraviolet divergences localized at the boundary, and the resulting boundary anomalous dimensions characterize the corresponding boundary universality class.


The boundary RG factors $Z_i^\partial$ are determined from the one-loop corrections to the boundary two-point functions $\langle\hat\psi\hat{\bar{\psi}}\rangle$ and $\langle\hat A_\mu\hat A_\mu\rangle$, with all external legs restricted to $z=0$, as illustrated in Fig.~\ref{fig:FeynmanDiagrams}(a,b). 
The boundary RG factor can be obtained as
\begin{eqnarray}
    Z_2^\partial = 1 - \frac{(4+w)e^2}{8\pi^2 \varepsilon} , \qquad Z_{3}^{\partial}=1-\frac{Ne^2}{4\pi^2\varepsilon} ,
\end{eqnarray}
where $w=1$ ($w=-1$) corresponds to Neumann (Dirichlet) boundary condition, respectively.
We defer the calculation details to the Supplemental Material. 
The corresponding boundary scaling dimensions of the fermion and gauge fields are $\Delta_2^\partial=\frac{d-1}{2}+\eta_2^\partial$ and $\Delta_3^\partial=\frac{d-2}{2}+\eta_3^\partial$, respectively, where $\eta_2^\partial=\frac12{\rm d}\log Z_2^\partial/{\rm d}\log\mu$ and $\eta_3^\partial=\frac12{\rm d}\log Z_3^\partial/{\rm d}\log\mu$. 
Table~\ref{tab:scaling-dim} summarizes the resulting boundary scaling dimensions for the Neumann and Dirichlet boundary conditions. 
These distinct scaling properties characterize the corresponding boundary universality classes.
As shown in Fig.~\ref{fig:FeynmanDiagrams}(a), the fermion self-energy contains the gauge propagator and therefore depends on its boundary condition, leading to different fermion scaling dimensions. In contrast, Fig.~\ref{fig:FeynmanDiagrams}(b) shows that the gauge-field renormalization is generated by the fermion polarization bubble, which is governed by the same reflecting fermion boundary condition in both cases, yielding the same gauge-field scaling dimension at one loop. 

{\it Composite operators.}---The boundary criticality discussed here presents a compelling opportunity for experimental investigation. 
For instance, in quantum spin liquids,  the spin fluctuation near the edge may be accessed by spin-polarized scanning tunneling microscopy~\cite{feldmeier2020local}. 
In the parton theory, these observables are represented by gauge invariant fermion bilinears. 
Among the fermion bilinears relevant to the present discussion, we focus on the scalar and pseudoscalar mass channels, $\mathcal{O}_S=\bar\psi_j\psi_j$ and $\mathcal{O}_P=\bar\psi_j\gamma^5\psi_j$, which acquire nontrivial anomalous dimensions from gauge fluctuations.
The fermion boundary condition and its Hermitian conjugate imply $\hat\psi_j=-\gamma^z\hat\psi_j$ and $\hat\psi_j=-\hat\psi_j^\dagger \gamma^z$. 
The boundary scaling channel satisfies $\hat{\mathcal{O}}_S=\hat{\bar\psi}_j\hat\psi_j=\hat\psi_j\gamma^z\gamma^0\gamma^z\hat\psi_j=-\hat{\bar{\psi}}_j\hat\psi_j=0$, and is forbidden by the fermion boundary condition.
To determine the anomalous dimension for the fermion bilinear $\mathcal O_P$, we evaluate the three-point function containing a boundary composite insertion, $\langle\hat\psi_j\hat{\mathcal{O}}_P\hat{\bar\psi}_j\rangle$, as illustrated in Fig.~\ref{fig:FeynmanDiagrams}. 
At the one-loop level, the correction takes the form
\begin{align}
    \int &\frac{{\rm d^{d-1}}q}{(2\pi)^{d-1}}{\rm d}z_1{\rm d}z_2G(p,0,z_1)\sum_{\mu,\nu}\gamma^\mu G(q,z_1,0)\gamma^5\nonumber\\
    &\times G(q,0,z_2)\gamma^\nu D_{\mu\nu}(p-q,z_1,z_2) G(p,z_2,0).
\end{align}
The integration over the momentum coordinates can be performed within spherical coordinates, while the remaining $z$-direction integral procudes the logarithmic divergence as approaching the boundary.
To characterize this contribution, we introduce the RG factor for boundary fermion bilinear $\hat{\mathcal{O}}_P=Z_{P} Z_{P}^\partial Z_2Z_2^\partial\hat{\mathcal{O}}_{P,R}$, 
where $Z_{\mathcal O_P}=Z_2Z_P=1+\frac{3e^2}{8\pi^2\varepsilon} $ is the RG factor from the bulk contribution. 
The integration leads to the RG factors \begin{equation}
    Z_{{\mathcal O}_P}^\partial\equiv Z_2^\partial Z_P^\partial=1+\frac{we^2}{8\pi^2\varepsilon},
\end{equation}
where $w=1$ ($w=-1$) corresponds to Neumann (Dirichlet) boundary condition, respectively.
We leave the calculation detail to the Supplemental Material.
The associated boundary anomalous dimension is given by, $\eta_{\mathcal{O}_P}^\partial={\rm d}\log Z_{\mathcal{O}_P}^\partial/{\rm d}\log\mu$, as summarized in Table~\ref{tab:scaling-dim}.

\begin{figure}[t]
\centering
\includegraphics[width=1.05\columnwidth]{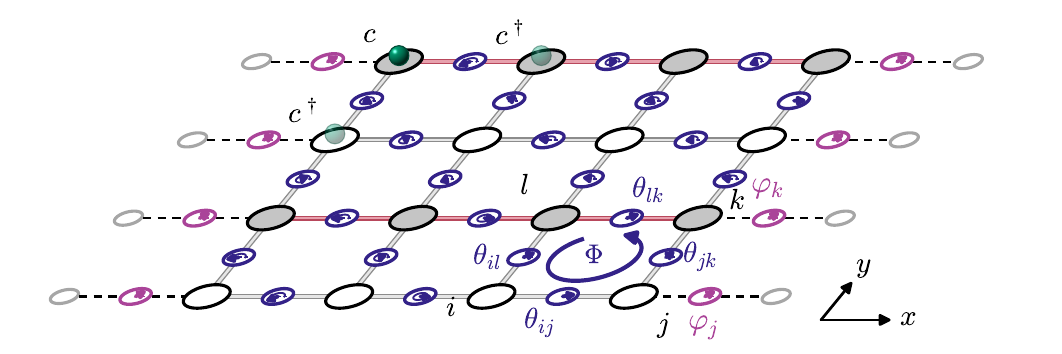}
\caption{Illustration of the fermion-gauge lattice model with $L=4$. 
The matter fields reside on the lattice, while the dynamical U(1) gauge fields reside on the solid bulk links.
The system has open boundary conditions along the $x$ direction and periodic boundary conditions along the $y$ axis.
Gray (white) sites denote $A$ ($B$) sublattices. 
The hopping amplitude between $A$ sublattice (pink link) is $-t$, and $t$ otherwise. The dashed virtual bonds contain the boundary-localized $U(1)$ phase field $\varphi_i$.}
\label{fig:piFluxLattice}
\end{figure}

\emph{Lattice model.}---A square-lattice model of fermions coupled to a compact $U(1)$ gauge boson provides a microscopic setting for the continuum theory.
The bulk Hamiltonian contains fermion hopping and gauge-field magnetic terms,
\begin{align}
    H=-t\sum_{\langle ij\rangle }\left({c_i}^\dagger e^{i\theta_{ij}}c_{j} + \text{H.c.} \right) +K\sum_{\square}\cos\left(\text{curl}(\theta)\right), 
\label{eq:lattice_hamiltonian}
\end{align}
where ${c_i}^\dagger$ ($c_i$) creates (annihilates) a fermion on site $i$, and $\theta_{ij}$ is a compact $U(1)$ gauge field defined on the oriented links $\langle ij\rangle$, with $\theta_{ij}=-\theta_{ji}$. 
The second term describes the dynamics of the gauge field, where $\sum_{\square}$ runs over all plaquettes, and $\text{curl}(\theta)=\sum_{ij\in \square}\theta_{ij}$ denotes the magnetic flux through each plaquette.
For $K>0$, the ground state favors a $\pi$-flux configuration on the square lattice, realizing a $U(1)$ deconfined spin liquid. 
The $\pi$-flux pattern doubles the unite cell and leads to two sublattices, labeled by $A$ and $B$. 
A convenient gauge choice is illustrated in Fig.~\ref{fig:piFluxLattice}, where the hopping amplitude is $-t$ on bonds (in pink color) within the $A$ sublattice and $+t$ on the remaining bonds.

In the absence of the boundaries, the $\pi$-flux phase hosts two distinct gapless Dirac cones located at $\mathbf{K}=\left(\frac{\pi}{2},\frac{\pi}{2}\right)$ and $\mathbf{K}'=\left(-\frac{\pi}{2},\frac{\pi}{2}\right)$. 
Adopting a $\mathbf{k}\cdot\mathbf{P}$ expansion around these two valleys and retaining only the linear term in momentum, we obtain the low-energy Dirac Hamiltonian, $\mathcal{H}=\tau^z\sigma^z k_x+\sigma^y k_y$, where $\tau$ ($\sigma$) acts in the valley (sublattice) space, respectively. 
The corresponding four-component spinor is defined as $\psi=(\psi_{A\K},\psi_{B\K},\psi_{A\K'},\psi_{B\K'})^T$. 
By introducing the $\gamma$ matrices $\gamma^\tau=\tau^y\sigma^z$, $\gamma^x=\tau^x$, and $\gamma^y=-\tau^y\sigma^x$, this Hamiltonian can be cast into Euclidean Dirac action in Eq.~\eqref{eq:fermion_action}.
To investigate boundary criticality, we introduce open boundaries at $x=0$ and $x=L$.
Note that we use $x$ to denote the boundary direction in the lattice model. 
These boundaries scatter the Dirac fermions and necessarily mix the two valleys. 
Requiring the fermion wave function to vanish outside the system leads to the boundary conditions
\begin{equation}
\begin{split}
& \psi_{\sigma\K}(0)+\psi_{\sigma\K}(0)=0, \\
& e^{i\frac{\pi}{2}L}\psi_{\sigma\K }(L)+e^{-i\frac{\pi}{2}L}\psi_{\sigma\K' }(L) =0,
\end{split}
\end{equation}
where $\sigma\in\{A,B\}$ labels the sublattice. 
These boundary conditions support gapless boundary modes for $L=2\mathbb{Z}$, corresponding to the conformal boundary condition for the fermion, $-\gamma^x\psi|_{x\in\text{bdy}}=\psi|_{x\in\text{bdy}}$.

By including the gauge fluctuation, the lattice formulation above provides a concrete ultraviolet-regularized model for studying boundary criticality in QED$_3$. 
The emergent $U(1)$ gauge fluctuation on top of a $\pi$-flux background is described in the low energy by an effective Hamiltonian $H_{\text{eff}}=K/2\int_{\mathcal{M}}{\rm d}^2x f_{xy}^2$ in the continuous limit, where $f_{xy}=\partial_x a_y-\partial_y a_x$ is the field strength. 
Opening a boundary in the $x$-direction, and varying the Hamiltonian leads to the boundary condition $f_{xy}\delta a_{y}|_{x\in\text{bdy}}=0$, where $f_{xy}|_{x\in\text{bdy}}=0$ corresponds to the Neumann boundary condition, whereas $\delta a_{y}|_{x\in\text{bdy}}=0$ realizes the Dirichlet boundary condition.
To realize these boundary conditions microscopically, we introduce a boundary interaction
\begin{align}
    H_{\text{{bdy}}}=-K_B\sum_{\langle ij\rangle\in \text{bdy}}\cos(\theta_{ij}-\varphi_i + \varphi_j),
\label{eq:bdy_hamiltonian}
\end{align}
where $\varphi_i$ is the charged field living on the boundary link as illustrated in Fig.~\ref{fig:piFluxLattice} and the sum runs over all boundary terms. 
Under a local $U(1)$ gauge transformation, $c_i\to e^{i\alpha_i}c_i$, $\theta_{ij}\to\theta_{ij}+\alpha_i-\alpha_j$, $\varphi_i\to\varphi_i+\alpha_i$, and hence the boundary action is gauge invariant.
When $K_B=0$, the boundary gauge field remains unconstrained, and the boundary equation of motion imposes the Neumann boundary condition. 
For $K_B\gg K$, the interaction locks $\theta_{ij}=\varphi_i-\varphi_j$ along the boundary, and pins the transverse component of the boundary field $a_{y}|_{x\in\text{bdy}}=\text{const}$.
This realizes the Dirichlet boundary condition.
Therefore, the Hamiltonian~\eqref{eq:lattice_hamiltonian} supplemented with the boundary terms~\eqref{eq:bdy_hamiltonian} provides a microscopic setting to realize the boundary criticality in QED$_3$.

A representative observable is the $(\pi,\pi)$ charge-density-wave (CDW) operator, denoted by $\mathcal O=\sum_{\boldsymbol{i}}(-1)^{i_x +i_y }c_{i}^\dagger c_{i}$.
In the continuous limit, the operator maps onto the fermion bilinear term $\psi^\dagger \tau^x\sigma^z\psi$, which is the pseudoscalar fermion bilinear $\hat{\mathcal O}_P = \hat{\bar\psi}\gamma^5\hat\psi$ on the boundary in our convention, with boundary dimension given in Table~\ref{tab:scaling-dim}. 
Hence, the boundary correlation of this CDW order reveals the boundary universality classes.

\emph{Concluding remarks.}---We have shown that the boundary condition of an emergent gauge field constitutes an independent universal datum of a $U(1)$ Dirac spin liquid. Using a half-space $4-\varepsilon$ expansion with reflecting Dirac fermions, we find that Neumann and Dirichlet gauge-field boundary conditions define distinct boundary universality classes. Notably, gauge-invariant edge correlations can therefore distinguish boundary states governed by the same bulk spin liquid.
We further introduced a fermion-gauge model in a square-lattice $\pi$-flux background that realizes these boundary conditions microscopically. 
Experimentally, spin-polarized tunneling and related local probes can access correlations near an edge~\cite{feldmeier2020local,konig2020tunneling,udagawa2021scanning,zhang2025edge}. Programmable Rydberg arrays provide a complementary platform in which boundary gauge couplings may be spatially controlled~\cite{Celi2019lqy}. These results establish gauge-field boundary conditions as a mechanism for generating distinct edge critical behavior in a fixed gapless quantum spin liquid and motivate nonperturbative studies of the resulting boundary phase structure.

\emph{Acknowledgements}---X.-J. Yu is supported by the National Natural Science Foundation of China (Grant No.12405034) and a start-up grant from Eastern Institute of Technology, Ningbo. 
The work of H. J. and S.-K.J. is supported by a start-up fund at Tulane University.
We used AI tools (GPT-5.6 and Claude Opus 5.5) to assist with figure generation and manuscript editing.

\emph{Note added.}---After the completion of this work, we became aware of Ref.~\cite{de2026conformal}, which studies conformal QED in $\mathrm{AdS}_d$ as a BCFT. 
Its results for the boundary free energy and singlet scalar operators complement our analysis of fermionic boundary operators and the microscopic realization of these boundary conditions in a fermion-gauge lattice model.

\bibliography{references.bib}

\begin{thebibliography}{83}%
\makeatletter
\providecommand \@ifxundefined [1]{%
 \@ifx{#1\undefined}
}%
\providecommand \@ifnum [1]{%
 \ifnum #1\expandafter \@firstoftwo
 \else \expandafter \@secondoftwo
 \fi
}%
\providecommand \@ifx [1]{%
 \ifx #1\expandafter \@firstoftwo
 \else \expandafter \@secondoftwo
 \fi
}%
\providecommand \natexlab [1]{#1}%
\providecommand \enquote  [1]{``#1''}%
\providecommand \bibnamefont  [1]{#1}%
\providecommand \bibfnamefont [1]{#1}%
\providecommand \citenamefont [1]{#1}%
\providecommand \href@noop [0]{\@secondoftwo}%
\providecommand \href [0]{\begingroup \@sanitize@url \@href}%
\providecommand \@href[1]{\@@startlink{#1}\@@href}%
\providecommand \@@href[1]{\endgroup#1\@@endlink}%
\providecommand \@sanitize@url [0]{\catcode `\\12\catcode `\$12\catcode
  `\&12\catcode `\#12\catcode `\^12\catcode `\_12\catcode `\%12\relax}%
\providecommand \@@startlink[1]{}%
\providecommand \@@endlink[0]{}%
\providecommand \url  [0]{\begingroup\@sanitize@url \@url }%
\providecommand \@url [1]{\endgroup\@href {#1}{\urlprefix }}%
\providecommand \urlprefix  [0]{URL }%
\providecommand \Eprint [0]{\href }%
\providecommand \doibase [0]{https://doi.org/}%
\providecommand \selectlanguage [0]{\@gobble}%
\providecommand \bibinfo  [0]{\@secondoftwo}%
\providecommand \bibfield  [0]{\@secondoftwo}%
\providecommand \translation [1]{[#1]}%
\providecommand \BibitemOpen [0]{}%
\providecommand \bibitemStop [0]{}%
\providecommand \bibitemNoStop [0]{.\EOS\space}%
\providecommand \EOS [0]{\spacefactor3000\relax}%
\providecommand \BibitemShut  [1]{\csname bibitem#1\endcsname}%
\let\auto@bib@innerbib\@empty
\bibitem [{\citenamefont {Andrei}\ \emph {et~al.}(2020)\citenamefont {Andrei},
  \citenamefont {Bissi}, \citenamefont {Buican}, \citenamefont {Cardy},
  \citenamefont {Dorey}, \citenamefont {Drukker}, \citenamefont {Erdmenger},
  \citenamefont {Friedan}, \citenamefont {Fursaev}, \citenamefont {Konechny}
  \emph {et~al.}}]{andrei2020boundary}%
  \BibitemOpen
  \bibfield  {author} {\bibinfo {author} {\bibfnamefont {N.}~\bibnamefont
  {Andrei}}, \bibinfo {author} {\bibfnamefont {A.}~\bibnamefont {Bissi}},
  \bibinfo {author} {\bibfnamefont {M.}~\bibnamefont {Buican}}, \bibinfo
  {author} {\bibfnamefont {J.}~\bibnamefont {Cardy}}, \bibinfo {author}
  {\bibfnamefont {P.}~\bibnamefont {Dorey}}, \bibinfo {author} {\bibfnamefont
  {N.}~\bibnamefont {Drukker}}, \bibinfo {author} {\bibfnamefont
  {J.}~\bibnamefont {Erdmenger}}, \bibinfo {author} {\bibfnamefont
  {D.}~\bibnamefont {Friedan}}, \bibinfo {author} {\bibfnamefont
  {D.}~\bibnamefont {Fursaev}}, \bibinfo {author} {\bibfnamefont
  {A.}~\bibnamefont {Konechny}}, \emph {et~al.},\ }\bibfield  {title} {\bibinfo
  {title} {Boundary and defect cft: open problems and applications},\
  }\href@noop {} {\bibfield  {journal} {\bibinfo  {journal} {Journal of Physics
  A: Mathematical and Theoretical}\ }\textbf {\bibinfo {volume} {53}},\
  \bibinfo {pages} {453002} (\bibinfo {year} {2020})}\BibitemShut {NoStop}%
\bibitem [{\citenamefont {Diehl}\ and\ \citenamefont
  {Dietrich}(1981)}]{diehl1981field}%
  \BibitemOpen
  \bibfield  {author} {\bibinfo {author} {\bibfnamefont {H.}~\bibnamefont
  {Diehl}}\ and\ \bibinfo {author} {\bibfnamefont {S.}~\bibnamefont
  {Dietrich}},\ }\bibfield  {title} {\bibinfo {title} {Field-theoretical
  approach to static critical phenomena in semi-infinite systems},\ }\href
  {https://link.springer.com/article/10.1007/BF01298293} {\bibfield  {journal}
  {\bibinfo  {journal} {Z. Phys. B}\ }\textbf {\bibinfo {volume} {42}},\
  \bibinfo {pages} {65} (\bibinfo {year} {1981})}\BibitemShut {NoStop}%
\bibitem [{\citenamefont {Diehl}\ and\ \citenamefont
  {Dietrich}(1983)}]{diehl1983multicritical}%
  \BibitemOpen
  \bibfield  {author} {\bibinfo {author} {\bibfnamefont {H.}~\bibnamefont
  {Diehl}}\ and\ \bibinfo {author} {\bibfnamefont {S.}~\bibnamefont
  {Dietrich}},\ }\bibfield  {title} {\bibinfo {title} {Multicritical behaviour
  at surfaces},\ }\href {https://link.springer.com/article/10.1007/BF01304094}
  {\bibfield  {journal} {\bibinfo  {journal} {Z. Phys. B}\ }\textbf {\bibinfo
  {volume} {50}},\ \bibinfo {pages} {117} (\bibinfo {year} {1983})}\BibitemShut
  {NoStop}%
\bibitem [{\citenamefont {Domb}\ and\ \citenamefont
  {Lebowitz}(1986)}]{domb1986phase}%
  \BibitemOpen
  \bibinfo {editor} {\bibfnamefont {C.}~\bibnamefont {Domb}}\ and\ \bibinfo
  {editor} {\bibfnamefont {J.~L.}\ \bibnamefont {Lebowitz}},\ eds.,\ \href@noop
  {} {\emph {\bibinfo {title} {Phase Transitions and Critical Phenomena}}},\
  Vol.~\bibinfo {volume} {10}\ (\bibinfo  {publisher} {Academic Press},\
  \bibinfo {address} {London},\ \bibinfo {year} {1986})\BibitemShut {NoStop}%
\bibitem [{\citenamefont {Diehl}(1997)}]{diehl1996the}%
  \BibitemOpen
  \bibfield  {author} {\bibinfo {author} {\bibfnamefont {H.~W.}\ \bibnamefont
  {Diehl}},\ }\bibfield  {title} {\bibinfo {title} {The theory of boundary
  critical phenomena},\ }\href
  {https://www.worldscientific.com/doi/abs/10.1142/S0217979297001751}
  {\bibfield  {journal} {\bibinfo  {journal} {Int. J. Mod. Phys. B}\ }\textbf
  {\bibinfo {volume} {11}},\ \bibinfo {pages} {3503} (\bibinfo {year}
  {1997})}\BibitemShut {NoStop}%
\bibitem [{\citenamefont {Cardy}(1996)}]{cardy1996scaling}%
  \BibitemOpen
  \bibfield  {author} {\bibinfo {author} {\bibfnamefont {J.}~\bibnamefont
  {Cardy}},\ }\href@noop {} {\emph {\bibinfo {title} {Scaling and
  renormalization in statistical physics}}},\ Vol.~\bibinfo {volume} {5}\
  (\bibinfo  {publisher} {Cambridge University Press},\ \bibinfo {address}
  {Cambridge, England},\ \bibinfo {year} {1996})\BibitemShut {NoStop}%
\bibitem [{\citenamefont {Liendo}\ \emph {et~al.}(2013)\citenamefont {Liendo},
  \citenamefont {Rastelli},\ and\ \citenamefont {van Rees}}]{liendo2012the}%
  \BibitemOpen
  \bibfield  {author} {\bibinfo {author} {\bibfnamefont {P.}~\bibnamefont
  {Liendo}}, \bibinfo {author} {\bibfnamefont {L.}~\bibnamefont {Rastelli}},\
  and\ \bibinfo {author} {\bibfnamefont {B.~C.}\ \bibnamefont {van Rees}},\
  }\bibfield  {title} {\bibinfo {title} {The bootstrap program for boundary
  cft$_d$},\ }\href {https://doi.org/10.1007/JHEP07(2013)113} {\bibfield
  {journal} {\bibinfo  {journal} {JHEP}\ }\textbf {\bibinfo {volume} {07}},\
  \bibinfo {pages} {113}}\BibitemShut {NoStop}%
\bibitem [{\citenamefont {Giombi}\ and\ \citenamefont
  {Khanchandani}(2020{\natexlab{a}})}]{giombi2020cft}%
  \BibitemOpen
  \bibfield  {author} {\bibinfo {author} {\bibfnamefont {S.}~\bibnamefont
  {Giombi}}\ and\ \bibinfo {author} {\bibfnamefont {H.}~\bibnamefont
  {Khanchandani}},\ }\bibfield  {title} {\bibinfo {title} {Cft in ads and
  boundary rg flows},\ }\href {https://doi.org/10.1007/JHEP11(2020)118}
  {\bibfield  {journal} {\bibinfo  {journal} {JHEP}\ }\textbf {\bibinfo
  {volume} {11}},\ \bibinfo {pages} {118}}\BibitemShut {NoStop}%
\bibitem [{\citenamefont {Giombi}\ and\ \citenamefont
  {Khanchandani}(2020{\natexlab{b}})}]{Giombi2020boundaryInteractions}%
  \BibitemOpen
  \bibfield  {author} {\bibinfo {author} {\bibfnamefont {S.}~\bibnamefont
  {Giombi}}\ and\ \bibinfo {author} {\bibfnamefont {H.}~\bibnamefont
  {Khanchandani}},\ }\bibfield  {title} {\bibinfo {title} {{$O(N)$} models with
  boundary interactions and their long-range generalizations},\ }\href
  {https://doi.org/10.1007/JHEP08(2020)010} {\bibfield  {journal} {\bibinfo
  {journal} {J. High Energy Phys.}\ }\textbf {\bibinfo {volume}
  {2020}}\bibfield  {number} {\bibinfo  {number} { (08)},\ \bibinfo {pages}
  {010}},\ }\Eprint {https://arxiv.org/abs/1912.08169} {arXiv:1912.08169
  [hep-th]} \BibitemShut {NoStop}%
\bibitem [{\citenamefont {Metlitski}(2022)}]{metlitski2020boundary}%
  \BibitemOpen
  \bibfield  {author} {\bibinfo {author} {\bibfnamefont {M.~A.}\ \bibnamefont
  {Metlitski}},\ }\bibfield  {title} {\bibinfo {title} {Boundary criticality of
  the o(n) model in $d=3$ critically revisited},\ }\href
  {https://www.scipost.org/10.21468/SciPostPhys.12.4.131?acad_field_slug=chemistry}
  {\bibfield  {journal} {\bibinfo  {journal} {SciPost Phys.}\ }\textbf
  {\bibinfo {volume} {12}},\ \bibinfo {pages} {131} (\bibinfo {year}
  {2022})}\BibitemShut {NoStop}%
\bibitem [{\citenamefont {Sun}\ and\ \citenamefont
  {Jian}(2025)}]{sun2025boundary}%
  \BibitemOpen
  \bibfield  {author} {\bibinfo {author} {\bibfnamefont {X.}~\bibnamefont
  {Sun}}\ and\ \bibinfo {author} {\bibfnamefont {S.-K.}\ \bibnamefont {Jian}},\
  }\bibfield  {title} {\bibinfo {title} {Boundary operator expansion and
  extraordinary phase transition in the tricritical o(n) model},\ }\href
  {https://doi.org/10.21468/SciPostPhys.18.6.210} {\bibfield  {journal}
  {\bibinfo  {journal} {SciPost Phys.}\ }\textbf {\bibinfo {volume} {18}},\
  \bibinfo {pages} {210} (\bibinfo {year} {2025})}\BibitemShut {NoStop}%
\bibitem [{\citenamefont {Sun}\ \emph {et~al.}(2026)\citenamefont {Sun},
  \citenamefont {Jian},\ and\ \citenamefont {Yao}}]{sun2026analytic}%
  \BibitemOpen
  \bibfield  {author} {\bibinfo {author} {\bibfnamefont {X.}~\bibnamefont
  {Sun}}, \bibinfo {author} {\bibfnamefont {S.-K.}\ \bibnamefont {Jian}},\ and\
  \bibinfo {author} {\bibfnamefont {H.}~\bibnamefont {Yao}},\ }\bibfield
  {title} {\bibinfo {title} {Analytic bootstrap for $ o (n) $ boundary
  conformal field theories with interacting boundaries},\ }\href@noop {}
  {\bibfield  {journal} {\bibinfo  {journal} {arXiv preprint arXiv:2605.28933}\
  } (\bibinfo {year} {2026})}\BibitemShut {NoStop}%
\bibitem [{\citenamefont {Giombi}\ \emph {et~al.}(2022)\citenamefont {Giombi},
  \citenamefont {Helfenberger},\ and\ \citenamefont
  {Khanchandani}}]{Giombi2021cnr}%
  \BibitemOpen
  \bibfield  {author} {\bibinfo {author} {\bibfnamefont {S.}~\bibnamefont
  {Giombi}}, \bibinfo {author} {\bibfnamefont {E.}~\bibnamefont
  {Helfenberger}},\ and\ \bibinfo {author} {\bibfnamefont {H.}~\bibnamefont
  {Khanchandani}},\ }\bibfield  {title} {\bibinfo {title} {Fermions in ads and
  gross-neveu bcft},\ }\href {https://doi.org/10.1007/JHEP07(2022)018}
  {\bibfield  {journal} {\bibinfo  {journal} {JHEP}\ }\textbf {\bibinfo
  {volume} {07}},\ \bibinfo {pages} {018}}\BibitemShut {NoStop}%
\bibitem [{\citenamefont {Herzog}\ and\ \citenamefont
  {Schaub}(2023)}]{Herzog2022bcb}%
  \BibitemOpen
  \bibfield  {author} {\bibinfo {author} {\bibfnamefont {C.~P.}\ \bibnamefont
  {Herzog}}\ and\ \bibinfo {author} {\bibfnamefont {V.}~\bibnamefont
  {Schaub}},\ }\bibfield  {title} {\bibinfo {title} {Fermions in boundary
  conformal field theory: crossing symmetry and e-expansion},\ }\href
  {https://doi.org/10.1007/JHEP02(2023)129} {\bibfield  {journal} {\bibinfo
  {journal} {JHEP}\ }\textbf {\bibinfo {volume} {02}},\ \bibinfo {pages}
  {129}}\BibitemShut {NoStop}%
\bibitem [{\citenamefont {Jiang}\ \emph {et~al.}(2025)\citenamefont {Jiang},
  \citenamefont {Ge},\ and\ \citenamefont {Jian}}]{jian2025bdygny}%
  \BibitemOpen
  \bibfield  {author} {\bibinfo {author} {\bibfnamefont {H.}~\bibnamefont
  {Jiang}}, \bibinfo {author} {\bibfnamefont {Y.}~\bibnamefont {Ge}},\ and\
  \bibinfo {author} {\bibfnamefont {S.-K.}\ \bibnamefont {Jian}},\ }\bibfield
  {title} {\bibinfo {title} {Boundary criticality for the gross-neveu-yukawa
  models},\ }\href {https://link.aps.org/doi/10.1103/tjfk-84f8} {\bibfield
  {journal} {\bibinfo  {journal} {Phys. Rev. Lett.}\ }\textbf {\bibinfo
  {volume} {135}},\ \bibinfo {pages} {141602} (\bibinfo {year}
  {2025})}\BibitemShut {NoStop}%
\bibitem [{\citenamefont {Fedorenko}\ and\ \citenamefont
  {Gruzberg}(2026)}]{Fedorenko2026bgny}%
  \BibitemOpen
  \bibfield  {author} {\bibinfo {author} {\bibfnamefont {A.~A.}\ \bibnamefont
  {Fedorenko}}\ and\ \bibinfo {author} {\bibfnamefont {I.~A.}\ \bibnamefont
  {Gruzberg}},\ }\bibfield  {title} {\bibinfo {title} {{Boundary critical
  behavior of the Gross-Neveu-Yukawa model}},\ }\href
  {https://doi.org/10.1103/2rm3-ggl3} {\bibfield  {journal} {\bibinfo
  {journal} {Phys. Rev. D}\ }\textbf {\bibinfo {volume} {114}},\ \bibinfo
  {pages} {025017} (\bibinfo {year} {2026})},\ \Eprint
  {https://arxiv.org/abs/2603.07637} {arXiv:2603.07637 [hep-th]} \BibitemShut
  {NoStop}%
\bibitem [{\citenamefont {Diatlyk}\ \emph {et~al.}(2026)\citenamefont
  {Diatlyk}, \citenamefont {Giombi},\ and\ \citenamefont
  {Sun}}]{Diatlyk2026higherOrderGNY}%
  \BibitemOpen
  \bibfield  {author} {\bibinfo {author} {\bibfnamefont {O.}~\bibnamefont
  {Diatlyk}}, \bibinfo {author} {\bibfnamefont {S.}~\bibnamefont {Giombi}},\
  and\ \bibinfo {author} {\bibfnamefont {Z.}~\bibnamefont {Sun}},\ }\bibfield
  {title} {\bibinfo {title} {Boundary criticality in the {Gross--Neveu--Yukawa}
  model at higher orders},\ }\bibfield  {journal} {\bibinfo  {journal} {arXiv
  e-prints}\ }\href {https://doi.org/10.48550/arXiv.2606.07510}
  {10.48550/arXiv.2606.07510} (\bibinfo {year} {2026}),\ \Eprint
  {https://arxiv.org/abs/2606.07510} {arXiv:2606.07510 [hep-th]} \BibitemShut
  {NoStop}%
\bibitem [{\citenamefont {Grover}\ \emph {et~al.}(2014)\citenamefont {Grover},
  \citenamefont {Sheng},\ and\ \citenamefont
  {Vishwanath}}]{grover2014emergent}%
  \BibitemOpen
  \bibfield  {author} {\bibinfo {author} {\bibfnamefont {T.}~\bibnamefont
  {Grover}}, \bibinfo {author} {\bibfnamefont {D.}~\bibnamefont {Sheng}},\ and\
  \bibinfo {author} {\bibfnamefont {A.}~\bibnamefont {Vishwanath}},\ }\bibfield
   {title} {\bibinfo {title} {Emergent space-time supersymmetry at the boundary
  of a topological phase},\ }\href@noop {} {\bibfield  {journal} {\bibinfo
  {journal} {Science}\ }\textbf {\bibinfo {volume} {344}},\ \bibinfo {pages}
  {280} (\bibinfo {year} {2014})}\BibitemShut {NoStop}%
\bibitem [{\citenamefont {Li}\ \emph {et~al.}(2017)\citenamefont {Li},
  \citenamefont {Jiang},\ and\ \citenamefont {Yao}}]{li2017edge}%
  \BibitemOpen
  \bibfield  {author} {\bibinfo {author} {\bibfnamefont {Z.-X.}\ \bibnamefont
  {Li}}, \bibinfo {author} {\bibfnamefont {Y.-F.}\ \bibnamefont {Jiang}},\ and\
  \bibinfo {author} {\bibfnamefont {H.}~\bibnamefont {Yao}},\ }\bibfield
  {title} {\bibinfo {title} {Edge quantum criticality and emergent
  supersymmetry in topological phases},\ }\href
  {https://doi.org/10.1103/PhysRevLett.119.107202} {\bibfield  {journal}
  {\bibinfo  {journal} {Phys. Rev. Lett.}\ }\textbf {\bibinfo {volume} {119}},\
  \bibinfo {pages} {107202} (\bibinfo {year} {2017})}\BibitemShut {NoStop}%
\bibitem [{\citenamefont {Ge}\ \emph {et~al.}(2026)\citenamefont {Ge},
  \citenamefont {Yao},\ and\ \citenamefont {Jian}}]{Ge2025ti}%
  \BibitemOpen
  \bibfield  {author} {\bibinfo {author} {\bibfnamefont {Y.}~\bibnamefont
  {Ge}}, \bibinfo {author} {\bibfnamefont {H.}~\bibnamefont {Yao}},\ and\
  \bibinfo {author} {\bibfnamefont {S.-K.}\ \bibnamefont {Jian}},\ }\bibfield
  {title} {\bibinfo {title} {Boundary criticality in two-dimensional
  interacting topological insulators},\ }\href
  {https://link.aps.org/doi/10.1103/ml88-hbd5} {\bibfield  {journal} {\bibinfo
  {journal} {Phys. Rev. B}\ }\textbf {\bibinfo {volume} {113}},\ \bibinfo
  {pages} {L121108} (\bibinfo {year} {2026})}\BibitemShut {NoStop}%
\bibitem [{\citenamefont {Ge}\ \emph {et~al.}(2025)\citenamefont {Ge},
  \citenamefont {Jiang}, \citenamefont {Yao},\ and\ \citenamefont
  {Jian}}]{Ge2025tsc}%
  \BibitemOpen
  \bibfield  {author} {\bibinfo {author} {\bibfnamefont {Y.}~\bibnamefont
  {Ge}}, \bibinfo {author} {\bibfnamefont {H.}~\bibnamefont {Jiang}}, \bibinfo
  {author} {\bibfnamefont {H.}~\bibnamefont {Yao}},\ and\ \bibinfo {author}
  {\bibfnamefont {S.-K.}\ \bibnamefont {Jian}},\ }\bibfield  {title} {\bibinfo
  {title} {Boundary criticality in two-dimensional correlated topological
  superconductors},\ }\href {https://arxiv.org/abs/2510.05230} {\bibfield
  {journal} {\bibinfo  {journal} {arXiv:2510.05230}\ } (\bibinfo {year}
  {2025})}\BibitemShut {NoStop}%
\bibitem [{\citenamefont {Shen}\ \emph {et~al.}(2025)\citenamefont {Shen},
  \citenamefont {Wu},\ and\ \citenamefont {Jian}}]{shen2025boundary}%
  \BibitemOpen
  \bibfield  {author} {\bibinfo {author} {\bibfnamefont {X.}~\bibnamefont
  {Shen}}, \bibinfo {author} {\bibfnamefont {Z.}~\bibnamefont {Wu}},\ and\
  \bibinfo {author} {\bibfnamefont {S.-K.}\ \bibnamefont {Jian}},\ }\bibfield
  {title} {\bibinfo {title} {Boundary and defect criticality in topological
  insulators and superconductors},\ }\href {https://doi.org/10.1103/4lv4-mc81}
  {\bibfield  {journal} {\bibinfo  {journal} {Phys. Rev. B}\ }\textbf {\bibinfo
  {volume} {112}},\ \bibinfo {pages} {L041118} (\bibinfo {year}
  {2025})}\BibitemShut {NoStop}%
\bibitem [{\citenamefont {Yu}\ \emph {et~al.}(2026{\natexlab{a}})\citenamefont
  {Yu}, \citenamefont {Xu},\ and\ \citenamefont {Lin}}]{yu2026topological}%
  \BibitemOpen
  \bibfield  {author} {\bibinfo {author} {\bibfnamefont {X.-J.}\ \bibnamefont
  {Yu}}, \bibinfo {author} {\bibfnamefont {L.}~\bibnamefont {Xu}},\ and\
  \bibinfo {author} {\bibfnamefont {H.-Q.}\ \bibnamefont {Lin}},\ }\bibfield
  {title} {\bibinfo {title} {Topological physics in quantum critical systems},\
  }\href {https://www.sciencedirect.com/science/article/pii/S0370157325002881}
  {\bibfield  {journal} {\bibinfo  {journal} {Physics Reports}\ }\textbf
  {\bibinfo {volume} {1160}},\ \bibinfo {pages} {1} (\bibinfo {year}
  {2026}{\natexlab{a}})}\BibitemShut {NoStop}%
\bibitem [{\citenamefont {Scaffidi}\ \emph {et~al.}(2017)\citenamefont
  {Scaffidi}, \citenamefont {Parker},\ and\ \citenamefont
  {Vasseur}}]{Scaffidi2017PRX}%
  \BibitemOpen
  \bibfield  {author} {\bibinfo {author} {\bibfnamefont {T.}~\bibnamefont
  {Scaffidi}}, \bibinfo {author} {\bibfnamefont {D.~E.}\ \bibnamefont
  {Parker}},\ and\ \bibinfo {author} {\bibfnamefont {R.}~\bibnamefont
  {Vasseur}},\ }\bibfield  {title} {\bibinfo {title} {Gapless
  symmetry-protected topological order},\ }\href
  {https://link.aps.org/doi/10.1103/PhysRevX.7.041048} {\bibfield  {journal}
  {\bibinfo  {journal} {Phys. Rev. X}\ }\textbf {\bibinfo {volume} {7}},\
  \bibinfo {pages} {041048} (\bibinfo {year} {2017})}\BibitemShut {NoStop}%
\bibitem [{\citenamefont {Verresen}\ \emph {et~al.}(2018)\citenamefont
  {Verresen}, \citenamefont {Jones},\ and\ \citenamefont
  {Pollmann}}]{Verresen2018PRL}%
  \BibitemOpen
  \bibfield  {author} {\bibinfo {author} {\bibfnamefont {R.}~\bibnamefont
  {Verresen}}, \bibinfo {author} {\bibfnamefont {N.~G.}\ \bibnamefont
  {Jones}},\ and\ \bibinfo {author} {\bibfnamefont {F.}~\bibnamefont
  {Pollmann}},\ }\bibfield  {title} {\bibinfo {title} {Topology and edge modes
  in quantum critical chains},\ }\href
  {https://link.aps.org/doi/10.1103/PhysRevLett.120.057001} {\bibfield
  {journal} {\bibinfo  {journal} {Phys. Rev. Lett.}\ }\textbf {\bibinfo
  {volume} {120}},\ \bibinfo {pages} {057001} (\bibinfo {year}
  {2018})}\BibitemShut {NoStop}%
\bibitem [{\citenamefont {Verresen}\ \emph {et~al.}(2021)\citenamefont
  {Verresen}, \citenamefont {Thorngren}, \citenamefont {Jones},\ and\
  \citenamefont {Pollmann}}]{Verresen2021PRX}%
  \BibitemOpen
  \bibfield  {author} {\bibinfo {author} {\bibfnamefont {R.}~\bibnamefont
  {Verresen}}, \bibinfo {author} {\bibfnamefont {R.}~\bibnamefont {Thorngren}},
  \bibinfo {author} {\bibfnamefont {N.~G.}\ \bibnamefont {Jones}},\ and\
  \bibinfo {author} {\bibfnamefont {F.}~\bibnamefont {Pollmann}},\ }\bibfield
  {title} {\bibinfo {title} {Gapless topological phases and symmetry-enriched
  quantum criticality},\ }\href
  {https://link.aps.org/doi/10.1103/PhysRevX.11.041059} {\bibfield  {journal}
  {\bibinfo  {journal} {Phys. Rev. X}\ }\textbf {\bibinfo {volume} {11}},\
  \bibinfo {pages} {041059} (\bibinfo {year} {2021})}\BibitemShut {NoStop}%
\bibitem [{\citenamefont {Thorngren}\ \emph {et~al.}(2021)\citenamefont
  {Thorngren}, \citenamefont {Vishwanath},\ and\ \citenamefont
  {Verresen}}]{Thorngren2021PRB}%
  \BibitemOpen
  \bibfield  {author} {\bibinfo {author} {\bibfnamefont {R.}~\bibnamefont
  {Thorngren}}, \bibinfo {author} {\bibfnamefont {A.}~\bibnamefont
  {Vishwanath}},\ and\ \bibinfo {author} {\bibfnamefont {R.}~\bibnamefont
  {Verresen}},\ }\bibfield  {title} {\bibinfo {title} {Intrinsically gapless
  topological phases},\ }\href
  {https://link.aps.org/doi/10.1103/PhysRevB.104.075132} {\bibfield  {journal}
  {\bibinfo  {journal} {Phys. Rev. B}\ }\textbf {\bibinfo {volume} {104}},\
  \bibinfo {pages} {075132} (\bibinfo {year} {2021})}\BibitemShut {NoStop}%
\bibitem [{\citenamefont {Duque}\ \emph {et~al.}(2021)\citenamefont {Duque},
  \citenamefont {Hu}, \citenamefont {You}, \citenamefont {Khemani},
  \citenamefont {Verresen},\ and\ \citenamefont {Vasseur}}]{Duque2021PRB}%
  \BibitemOpen
  \bibfield  {author} {\bibinfo {author} {\bibfnamefont {C.~M.}\ \bibnamefont
  {Duque}}, \bibinfo {author} {\bibfnamefont {H.-Y.}\ \bibnamefont {Hu}},
  \bibinfo {author} {\bibfnamefont {Y.-Z.}\ \bibnamefont {You}}, \bibinfo
  {author} {\bibfnamefont {V.}~\bibnamefont {Khemani}}, \bibinfo {author}
  {\bibfnamefont {R.}~\bibnamefont {Verresen}},\ and\ \bibinfo {author}
  {\bibfnamefont {R.}~\bibnamefont {Vasseur}},\ }\bibfield  {title} {\bibinfo
  {title} {Topological and symmetry-enriched random quantum critical points},\
  }\href {https://link.aps.org/doi/10.1103/PhysRevB.103.L100207} {\bibfield
  {journal} {\bibinfo  {journal} {Phys. Rev. B}\ }\textbf {\bibinfo {volume}
  {103}},\ \bibinfo {pages} {L100207} (\bibinfo {year} {2021})}\BibitemShut
  {NoStop}%
\bibitem [{\citenamefont {Yu}\ \emph {et~al.}(2022)\citenamefont {Yu},
  \citenamefont {Huang}, \citenamefont {Song}, \citenamefont {Xu},
  \citenamefont {Ding},\ and\ \citenamefont {Zhang}}]{Yu2022PRL}%
  \BibitemOpen
  \bibfield  {author} {\bibinfo {author} {\bibfnamefont {X.-J.}\ \bibnamefont
  {Yu}}, \bibinfo {author} {\bibfnamefont {R.-Z.}\ \bibnamefont {Huang}},
  \bibinfo {author} {\bibfnamefont {H.-H.}\ \bibnamefont {Song}}, \bibinfo
  {author} {\bibfnamefont {L.}~\bibnamefont {Xu}}, \bibinfo {author}
  {\bibfnamefont {C.}~\bibnamefont {Ding}},\ and\ \bibinfo {author}
  {\bibfnamefont {L.}~\bibnamefont {Zhang}},\ }\bibfield  {title} {\bibinfo
  {title} {Conformal boundary conditions of symmetry-enriched quantum critical
  spin chains},\ }\href {https://doi.org/10.1103/PhysRevLett.129.210601}
  {\bibfield  {journal} {\bibinfo  {journal} {Phys. Rev. Lett.}\ }\textbf
  {\bibinfo {volume} {129}},\ \bibinfo {pages} {210601} (\bibinfo {year}
  {2022})}\BibitemShut {NoStop}%
\bibitem [{\citenamefont {Yu}\ \emph {et~al.}(2024)\citenamefont {Yu},
  \citenamefont {Yang}, \citenamefont {Lin},\ and\ \citenamefont
  {Jian}}]{Yu2024PRL}%
  \BibitemOpen
  \bibfield  {author} {\bibinfo {author} {\bibfnamefont {X.-J.}\ \bibnamefont
  {Yu}}, \bibinfo {author} {\bibfnamefont {S.}~\bibnamefont {Yang}}, \bibinfo
  {author} {\bibfnamefont {H.-Q.}\ \bibnamefont {Lin}},\ and\ \bibinfo {author}
  {\bibfnamefont {S.-K.}\ \bibnamefont {Jian}},\ }\bibfield  {title} {\bibinfo
  {title} {Universal entanglement spectrum in one-dimensional gapless symmetry
  protected topological states},\ }\href
  {https://doi.org/10.1103/PhysRevLett.133.026601} {\bibfield  {journal}
  {\bibinfo  {journal} {Phys. Rev. Lett.}\ }\textbf {\bibinfo {volume} {133}},\
  \bibinfo {pages} {026601} (\bibinfo {year} {2024})}\BibitemShut {NoStop}%
\bibitem [{\citenamefont {Prembabu}\ \emph {et~al.}(2024)\citenamefont
  {Prembabu}, \citenamefont {Thorngren},\ and\ \citenamefont
  {Verresen}}]{Prembabu2024PRB}%
  \BibitemOpen
  \bibfield  {author} {\bibinfo {author} {\bibfnamefont {S.}~\bibnamefont
  {Prembabu}}, \bibinfo {author} {\bibfnamefont {R.}~\bibnamefont
  {Thorngren}},\ and\ \bibinfo {author} {\bibfnamefont {R.}~\bibnamefont
  {Verresen}},\ }\bibfield  {title} {\bibinfo {title} {Boundary-deconfined
  quantum criticality at transitions between symmetry-protected topological
  chains},\ }\href {https://doi.org/10.1103/PhysRevB.109.L201112} {\bibfield
  {journal} {\bibinfo  {journal} {Phys. Rev. B}\ }\textbf {\bibinfo {volume}
  {109}},\ \bibinfo {pages} {L201112} (\bibinfo {year} {2024})}\BibitemShut
  {NoStop}%
\bibitem [{\citenamefont {Yu}\ and\ \citenamefont {Li}(2024)}]{Yu2024PRB}%
  \BibitemOpen
  \bibfield  {author} {\bibinfo {author} {\bibfnamefont {X.-J.}\ \bibnamefont
  {Yu}}\ and\ \bibinfo {author} {\bibfnamefont {W.-L.}\ \bibnamefont {Li}},\
  }\bibfield  {title} {\bibinfo {title} {Fidelity susceptibility at the
  lifshitz transition between the noninteracting topologically distinct quantum
  critical points},\ }\href {https://doi.org/10.1103/PhysRevB.110.045119}
  {\bibfield  {journal} {\bibinfo  {journal} {Phys. Rev. B}\ }\textbf {\bibinfo
  {volume} {110}},\ \bibinfo {pages} {045119} (\bibinfo {year}
  {2024})}\BibitemShut {NoStop}%
\bibitem [{\citenamefont {Zhang}\ \emph {et~al.}(2024)\citenamefont {Zhang},
  \citenamefont {Li}, \citenamefont {Yang},\ and\ \citenamefont
  {Yu}}]{Zhang2024PRA}%
  \BibitemOpen
  \bibfield  {author} {\bibinfo {author} {\bibfnamefont {H.-L.}\ \bibnamefont
  {Zhang}}, \bibinfo {author} {\bibfnamefont {H.-Z.}\ \bibnamefont {Li}},
  \bibinfo {author} {\bibfnamefont {S.}~\bibnamefont {Yang}},\ and\ \bibinfo
  {author} {\bibfnamefont {X.-J.}\ \bibnamefont {Yu}},\ }\bibfield  {title}
  {\bibinfo {title} {Quantum phase transition and critical behavior between the
  gapless topological phases},\ }\href
  {https://doi.org/10.1103/PhysRevA.109.062226} {\bibfield  {journal} {\bibinfo
   {journal} {Phys. Rev. A}\ }\textbf {\bibinfo {volume} {109}},\ \bibinfo
  {pages} {062226} (\bibinfo {year} {2024})}\BibitemShut {NoStop}%
\bibitem [{\citenamefont {Li}\ \emph {et~al.}(2025)\citenamefont {Li},
  \citenamefont {Oshikawa},\ and\ \citenamefont {Zheng}}]{Li2025SciPost}%
  \BibitemOpen
  \bibfield  {author} {\bibinfo {author} {\bibfnamefont {L.}~\bibnamefont
  {Li}}, \bibinfo {author} {\bibfnamefont {M.}~\bibnamefont {Oshikawa}},\ and\
  \bibinfo {author} {\bibfnamefont {Y.}~\bibnamefont {Zheng}},\ }\bibfield
  {title} {\bibinfo {title} {Intrinsically/purely gapless-spt from
  non-invertible duality transformations},\ }\href
  {https://doi.org/10.21468/SciPostPhys.18.5.153} {\bibfield  {journal}
  {\bibinfo  {journal} {SciPost Phys.}\ }\textbf {\bibinfo {volume} {18}},\
  \bibinfo {pages} {153} (\bibinfo {year} {2025})}\BibitemShut {NoStop}%
\bibitem [{\citenamefont {Huang}\ and\ \citenamefont
  {Cheng}(2025)}]{Huang2025SciPost}%
  \BibitemOpen
  \bibfield  {author} {\bibinfo {author} {\bibfnamefont {S.-J.}\ \bibnamefont
  {Huang}}\ and\ \bibinfo {author} {\bibfnamefont {M.}~\bibnamefont {Cheng}},\
  }\bibfield  {title} {\bibinfo {title} {Topological holography, quantum
  criticality, and boundary states},\ }\href
  {https://doi.org/10.21468/SciPostPhys.18.6.213} {\bibfield  {journal}
  {\bibinfo  {journal} {SciPost Phys.}\ }\textbf {\bibinfo {volume} {18}},\
  \bibinfo {pages} {213} (\bibinfo {year} {2025})}\BibitemShut {NoStop}%
\bibitem [{\citenamefont {Deng}\ \emph {et~al.}(2026)\citenamefont {Deng},
  \citenamefont {Yang}, \citenamefont {Sun}, \citenamefont {Li},\ and\
  \citenamefont {Yu}}]{Deng2026PRL}%
  \BibitemOpen
  \bibfield  {author} {\bibinfo {author} {\bibfnamefont {M.}~\bibnamefont
  {Deng}}, \bibinfo {author} {\bibfnamefont {S.}~\bibnamefont {Yang}}, \bibinfo
  {author} {\bibfnamefont {C.}~\bibnamefont {Sun}}, \bibinfo {author}
  {\bibfnamefont {F.}~\bibnamefont {Li}},\ and\ \bibinfo {author}
  {\bibfnamefont {X.-J.}\ \bibnamefont {Yu}},\ }\bibfield  {title} {\bibinfo
  {title} {Anomalous dynamical scaling at topological quantum criticality},\
  }\href {https://doi.org/10.1103/mqr4-wnny} {\bibfield  {journal} {\bibinfo
  {journal} {Phys. Rev. Lett.}\ }\textbf {\bibinfo {volume} {137}},\ \bibinfo
  {pages} {096605} (\bibinfo {year} {2026})}\BibitemShut {NoStop}%
\bibitem [{\citenamefont {Zhong}\ \emph {et~al.}(2025)\citenamefont {Zhong},
  \citenamefont {Lin},\ and\ \citenamefont {Yu}}]{Zhong2025PRB}%
  \BibitemOpen
  \bibfield  {author} {\bibinfo {author} {\bibfnamefont {W.-H.}\ \bibnamefont
  {Zhong}}, \bibinfo {author} {\bibfnamefont {H.-Q.}\ \bibnamefont {Lin}},\
  and\ \bibinfo {author} {\bibfnamefont {X.-J.}\ \bibnamefont {Yu}},\
  }\bibfield  {title} {\bibinfo {title} {Quantum entanglement of fermionic
  symmetry-enriched quantum critical points in one dimension},\ }\href
  {https://doi.org/10.1103/cv5q-8t25} {\bibfield  {journal} {\bibinfo
  {journal} {Phys. Rev. B}\ }\textbf {\bibinfo {volume} {112}},\ \bibinfo
  {pages} {075129} (\bibinfo {year} {2025})}\BibitemShut {NoStop}%
\bibitem [{\citenamefont {Yang}\ \emph
  {et~al.}(2026{\natexlab{a}})\citenamefont {Yang}, \citenamefont {Xu},
  \citenamefont {Lu}, \citenamefont {You}, \citenamefont {Lin},\ and\
  \citenamefont {Yu}}]{Yang2026PRBL}%
  \BibitemOpen
  \bibfield  {author} {\bibinfo {author} {\bibfnamefont {S.}~\bibnamefont
  {Yang}}, \bibinfo {author} {\bibfnamefont {F.}~\bibnamefont {Xu}}, \bibinfo
  {author} {\bibfnamefont {D.-C.}\ \bibnamefont {Lu}}, \bibinfo {author}
  {\bibfnamefont {Y.-Z.}\ \bibnamefont {You}}, \bibinfo {author} {\bibfnamefont
  {H.-Q.}\ \bibnamefont {Lin}},\ and\ \bibinfo {author} {\bibfnamefont {X.-J.}\
  \bibnamefont {Yu}},\ }\bibfield  {title} {\bibinfo {title} {Deconfined
  criticality as intrinsically gapless topological state in one dimension},\
  }\href {https://doi.org/10.1103/nj3d-8g9s} {\bibfield  {journal} {\bibinfo
  {journal} {Phys. Rev. B}\ }\textbf {\bibinfo {volume} {113}},\ \bibinfo
  {pages} {L201105} (\bibinfo {year} {2026}{\natexlab{a}})}\BibitemShut
  {NoStop}%
\bibitem [{\citenamefont {Guo}\ \emph {et~al.}(2026)\citenamefont {Guo},
  \citenamefont {Yang},\ and\ \citenamefont {Yu}}]{guo2026lihaldane}%
  \BibitemOpen
  \bibfield  {author} {\bibinfo {author} {\bibfnamefont {Y.}~\bibnamefont
  {Guo}}, \bibinfo {author} {\bibfnamefont {S.}~\bibnamefont {Yang}},\ and\
  \bibinfo {author} {\bibfnamefont {X.-J.}\ \bibnamefont {Yu}},\ }\bibfield
  {title} {\bibinfo {title} {Generalized li-haldane correspondence in critical
  free-fermion systems},\ }\href {https://doi.org/10.1103/96gp-fq3j} {\bibfield
   {journal} {\bibinfo  {journal} {Phys. Rev. Res.}\ }\textbf {\bibinfo
  {volume} {8}},\ \bibinfo {pages} {023203} (\bibinfo {year}
  {2026})}\BibitemShut {NoStop}%
\bibitem [{\citenamefont {Prembabu}\ and\ \citenamefont
  {Verresen}(2025)}]{prembabu2025multi}%
  \BibitemOpen
  \bibfield  {author} {\bibinfo {author} {\bibfnamefont {S.}~\bibnamefont
  {Prembabu}}\ and\ \bibinfo {author} {\bibfnamefont {R.}~\bibnamefont
  {Verresen}},\ }\bibfield  {title} {\bibinfo {title} {Multicriticality between
  purely gapless spt phases with unitary symmetry},\ }\href
  {https://arxiv.org/abs/2509.20431} {\bibfield  {journal} {\bibinfo  {journal}
  {arXiv:2509.20431}\ } (\bibinfo {year} {2025})}\BibitemShut {NoStop}%
\bibitem [{\citenamefont {Prembabu}\ \emph {et~al.}(2025)\citenamefont
  {Prembabu}, \citenamefont {Shao},\ and\ \citenamefont
  {Verresen}}]{prembabu2025non}%
  \BibitemOpen
  \bibfield  {author} {\bibinfo {author} {\bibfnamefont {S.}~\bibnamefont
  {Prembabu}}, \bibinfo {author} {\bibfnamefont {S.-H.}\ \bibnamefont {Shao}},\
  and\ \bibinfo {author} {\bibfnamefont {R.}~\bibnamefont {Verresen}},\
  }\bibfield  {title} {\bibinfo {title} {Non-invertible interfaces between
  symmetry-enriched critical phases},\ }\href
  {https://arxiv.org/abs/2512.23706} {\bibfield  {journal} {\bibinfo  {journal}
  {arXiv:2512.23706}\ } (\bibinfo {year} {2025})}\BibitemShut {NoStop}%
\bibitem [{\citenamefont {Yu}\ \emph {et~al.}(2026{\natexlab{b}})\citenamefont
  {Yu}, \citenamefont {Yang}, \citenamefont {Liu}, \citenamefont {Lin},\ and\
  \citenamefont {Jian}}]{Yu2026PRB}%
  \BibitemOpen
  \bibfield  {author} {\bibinfo {author} {\bibfnamefont {X.-J.}\ \bibnamefont
  {Yu}}, \bibinfo {author} {\bibfnamefont {S.}~\bibnamefont {Yang}}, \bibinfo
  {author} {\bibfnamefont {S.}~\bibnamefont {Liu}}, \bibinfo {author}
  {\bibfnamefont {H.-Q.}\ \bibnamefont {Lin}},\ and\ \bibinfo {author}
  {\bibfnamefont {S.-K.}\ \bibnamefont {Jian}},\ }\bibfield  {title} {\bibinfo
  {title} {Gapless symmetry-protected topological states in measurement-only
  circuits},\ }\href {https://doi.org/10.1103/b95c-th5t} {\bibfield  {journal}
  {\bibinfo  {journal} {Phys. Rev. B}\ }\textbf {\bibinfo {volume} {113}},\
  \bibinfo {pages} {134302} (\bibinfo {year} {2026}{\natexlab{b}})}\BibitemShut
  {NoStop}%
\bibitem [{\citenamefont {Tan}\ \emph {et~al.}(2026)\citenamefont {Tan},
  \citenamefont {Wang}, \citenamefont {Yang}, \citenamefont {Shen},
  \citenamefont {Jin}, \citenamefont {Zhu}, \citenamefont {Ji}, \citenamefont
  {Xu}, \citenamefont {Chen}, \citenamefont {Wu}, \citenamefont {Zhang},
  \citenamefont {Gao}, \citenamefont {Wang}, \citenamefont {Zou}, \citenamefont
  {Zhang}, \citenamefont {Li}, \citenamefont {Bao}, \citenamefont {Zhu},
  \citenamefont {Zhong}, \citenamefont {Cui}, \citenamefont {Han},
  \citenamefont {He}, \citenamefont {Wang}, \citenamefont {Yang}, \citenamefont
  {Wang}, \citenamefont {Shen}, \citenamefont {Liu}, \citenamefont {Song},
  \citenamefont {Deng}, \citenamefont {Dong}, \citenamefont {Zhang},
  \citenamefont {Jian}, \citenamefont {Li}, \citenamefont {Wang}, \citenamefont
  {Zhang}, \citenamefont {Guo}, \citenamefont {Lin}, \citenamefont {Song},
  \citenamefont {Yu}, \citenamefont {Wang},\ and\ \citenamefont
  {Wu}}]{Tan2026CP}%
  \BibitemOpen
  \bibfield  {author} {\bibinfo {author} {\bibfnamefont {Z.}~\bibnamefont
  {Tan}}, \bibinfo {author} {\bibfnamefont {K.}~\bibnamefont {Wang}}, \bibinfo
  {author} {\bibfnamefont {S.}~\bibnamefont {Yang}}, \bibinfo {author}
  {\bibfnamefont {F.}~\bibnamefont {Shen}}, \bibinfo {author} {\bibfnamefont
  {F.}~\bibnamefont {Jin}}, \bibinfo {author} {\bibfnamefont {X.}~\bibnamefont
  {Zhu}}, \bibinfo {author} {\bibfnamefont {Y.}~\bibnamefont {Ji}}, \bibinfo
  {author} {\bibfnamefont {S.}~\bibnamefont {Xu}}, \bibinfo {author}
  {\bibfnamefont {J.}~\bibnamefont {Chen}}, \bibinfo {author} {\bibfnamefont
  {Y.}~\bibnamefont {Wu}}, \bibinfo {author} {\bibfnamefont {C.}~\bibnamefont
  {Zhang}}, \bibinfo {author} {\bibfnamefont {Y.}~\bibnamefont {Gao}}, \bibinfo
  {author} {\bibfnamefont {N.}~\bibnamefont {Wang}}, \bibinfo {author}
  {\bibfnamefont {Y.}~\bibnamefont {Zou}}, \bibinfo {author} {\bibfnamefont
  {A.}~\bibnamefont {Zhang}}, \bibinfo {author} {\bibfnamefont
  {T.}~\bibnamefont {Li}}, \bibinfo {author} {\bibfnamefont {Z.}~\bibnamefont
  {Bao}}, \bibinfo {author} {\bibfnamefont {Z.}~\bibnamefont {Zhu}}, \bibinfo
  {author} {\bibfnamefont {J.}~\bibnamefont {Zhong}}, \bibinfo {author}
  {\bibfnamefont {Z.}~\bibnamefont {Cui}}, \bibinfo {author} {\bibfnamefont
  {Y.}~\bibnamefont {Han}}, \bibinfo {author} {\bibfnamefont {Y.}~\bibnamefont
  {He}}, \bibinfo {author} {\bibfnamefont {H.}~\bibnamefont {Wang}}, \bibinfo
  {author} {\bibfnamefont {J.}~\bibnamefont {Yang}}, \bibinfo {author}
  {\bibfnamefont {Y.}~\bibnamefont {Wang}}, \bibinfo {author} {\bibfnamefont
  {J.}~\bibnamefont {Shen}}, \bibinfo {author} {\bibfnamefont {G.}~\bibnamefont
  {Liu}}, \bibinfo {author} {\bibfnamefont {Z.}~\bibnamefont {Song}}, \bibinfo
  {author} {\bibfnamefont {J.}~\bibnamefont {Deng}}, \bibinfo {author}
  {\bibfnamefont {H.}~\bibnamefont {Dong}}, \bibinfo {author} {\bibfnamefont
  {P.}~\bibnamefont {Zhang}}, \bibinfo {author} {\bibfnamefont {S.-K.}\
  \bibnamefont {Jian}}, \bibinfo {author} {\bibfnamefont {H.}~\bibnamefont
  {Li}}, \bibinfo {author} {\bibfnamefont {Z.}~\bibnamefont {Wang}}, \bibinfo
  {author} {\bibfnamefont {J.}~\bibnamefont {Zhang}}, \bibinfo {author}
  {\bibfnamefont {Q.}~\bibnamefont {Guo}}, \bibinfo {author} {\bibfnamefont
  {H.-Q.}\ \bibnamefont {Lin}}, \bibinfo {author} {\bibfnamefont
  {C.}~\bibnamefont {Song}}, \bibinfo {author} {\bibfnamefont {X.-J.}\
  \bibnamefont {Yu}}, \bibinfo {author} {\bibfnamefont {H.}~\bibnamefont
  {Wang}},\ and\ \bibinfo {author} {\bibfnamefont {F.}~\bibnamefont {Wu}},\
  }\bibfield  {title} {\bibinfo {title} {Exploring nontrivial topology at
  quantum criticality on a superconducting processor},\ }\href
  {https://doi.org/10.1038/s42005-026-02569-9} {\bibfield  {journal} {\bibinfo
  {journal} {Communications Physics}\ }\textbf {\bibinfo {volume} {9}},\
  \bibinfo {pages} {136} (\bibinfo {year} {2026})}\BibitemShut {NoStop}%
\bibitem [{\citenamefont {Xu}\ \emph {et~al.}(2026)\citenamefont {Xu},
  \citenamefont {Pollmann},\ and\ \citenamefont {Knap}}]{xu2026frame}%
  \BibitemOpen
  \bibfield  {author} {\bibinfo {author} {\bibfnamefont {W.-T.}\ \bibnamefont
  {Xu}}, \bibinfo {author} {\bibfnamefont {F.}~\bibnamefont {Pollmann}},\ and\
  \bibinfo {author} {\bibfnamefont {M.}~\bibnamefont {Knap}},\ }\href
  {https://arxiv.org/abs/2604.10128} {\bibinfo {title} {A framework for
  predicting entanglement spectra of gapless symmetry-protected topological
  states in one dimension}} (\bibinfo {year} {2026}),\ \Eprint
  {https://arxiv.org/abs/2604.10128} {arXiv:2604.10128 [quant-ph]} \BibitemShut
  {NoStop}%
\bibitem [{\citenamefont {Yang}\ \emph
  {et~al.}(2026{\natexlab{b}})\citenamefont {Yang}, \citenamefont {Lin},\ and\
  \citenamefont {Yu}}]{yang2026topologicaltricriticalisinguniversality}%
  \BibitemOpen
  \bibfield  {author} {\bibinfo {author} {\bibfnamefont {S.}~\bibnamefont
  {Yang}}, \bibinfo {author} {\bibfnamefont {H.-Q.}\ \bibnamefont {Lin}},\ and\
  \bibinfo {author} {\bibfnamefont {X.-J.}\ \bibnamefont {Yu}},\ }\href
  {https://arxiv.org/abs/2606.15588} {\bibinfo {title} {Topological tricritical
  ising universality class in one dimension}} (\bibinfo {year}
  {2026}{\natexlab{b}}),\ \Eprint {https://arxiv.org/abs/2606.15588}
  {arXiv:2606.15588 [cond-mat.str-el]} \BibitemShut {NoStop}%
\bibitem [{\citenamefont {Chou}\ and\ \citenamefont
  {Yu}(2026)}]{chou2026topologicallyenforcedlifshitzmulticriticality}%
  \BibitemOpen
  \bibfield  {author} {\bibinfo {author} {\bibfnamefont {K.-H.}\ \bibnamefont
  {Chou}}\ and\ \bibinfo {author} {\bibfnamefont {X.-J.}\ \bibnamefont {Yu}},\
  }\href {https://arxiv.org/abs/2606.07380} {\bibinfo {title} {Topologically
  enforced lifshitz multicriticality in one dimension}} (\bibinfo {year}
  {2026}),\ \Eprint {https://arxiv.org/abs/2606.07380} {arXiv:2606.07380
  [cond-mat.stat-mech]} \BibitemShut {NoStop}%
\bibitem [{\citenamefont {Chou}\ \emph {et~al.}(2026)\citenamefont {Chou},
  \citenamefont {Yu},\ and\ \citenamefont
  {Chang}}]{chou2026ptsymmetryenrichednonunitarycriticality}%
  \BibitemOpen
  \bibfield  {author} {\bibinfo {author} {\bibfnamefont {K.-H.}\ \bibnamefont
  {Chou}}, \bibinfo {author} {\bibfnamefont {X.-J.}\ \bibnamefont {Yu}},\ and\
  \bibinfo {author} {\bibfnamefont {P.-Y.}\ \bibnamefont {Chang}},\ }\href
  {https://arxiv.org/abs/2509.09587} {\bibinfo {title} {Pt symmetry-enriched
  non-unitary criticality}} (\bibinfo {year} {2026}),\ \Eprint
  {https://arxiv.org/abs/2509.09587} {arXiv:2509.09587 [quant-ph]} \BibitemShut
  {NoStop}%
\bibitem [{\citenamefont {Chen}\ \emph {et~al.}(2026)\citenamefont {Chen},
  \citenamefont {Jing}, \citenamefont {Nori},\ and\ \citenamefont
  {Yu}}]{chen2026criticaltopologicalphotonicssynthetic}%
  \BibitemOpen
  \bibfield  {author} {\bibinfo {author} {\bibfnamefont {M.}~\bibnamefont
  {Chen}}, \bibinfo {author} {\bibfnamefont {Y.}~\bibnamefont {Jing}}, \bibinfo
  {author} {\bibfnamefont {F.}~\bibnamefont {Nori}},\ and\ \bibinfo {author}
  {\bibfnamefont {X.-J.}\ \bibnamefont {Yu}},\ }\href
  {https://arxiv.org/abs/2608.21791} {\bibinfo {title} {Critical topological
  photonics in synthetic dimensions}} (\bibinfo {year} {2026}),\ \Eprint
  {https://arxiv.org/abs/2608.21791} {arXiv:2608.21791 [physics.optics]}
  \BibitemShut {NoStop}%
\bibitem [{\citenamefont {Savary}\ and\ \citenamefont
  {Balents}(2017)}]{savary2016quantum}%
  \BibitemOpen
  \bibfield  {author} {\bibinfo {author} {\bibfnamefont {L.}~\bibnamefont
  {Savary}}\ and\ \bibinfo {author} {\bibfnamefont {L.}~\bibnamefont
  {Balents}},\ }\bibfield  {title} {\bibinfo {title} {Quantum spin liquids: a
  review},\ }\href {https://doi.org/10.1088/0034-4885/80/1/016502} {\bibfield
  {journal} {\bibinfo  {journal} {Reports on Progress in Physics}\ }\textbf
  {\bibinfo {volume} {80}},\ \bibinfo {pages} {016502} (\bibinfo {year}
  {2017})}\BibitemShut {NoStop}%
\bibitem [{\citenamefont {Rantner}\ and\ \citenamefont
  {Wen}(2002)}]{wen2002aQSL}%
  \BibitemOpen
  \bibfield  {author} {\bibinfo {author} {\bibfnamefont {W.}~\bibnamefont
  {Rantner}}\ and\ \bibinfo {author} {\bibfnamefont {X.-G.}\ \bibnamefont
  {Wen}},\ }\bibfield  {title} {\bibinfo {title} {Spin correlations in the
  algebraic spin liquid: Implications for high-tc superconductors},\ }\href
  {https://journals.aps.org/prb/abstract/10.1103/PhysRevB.66.144501} {\bibfield
   {journal} {\bibinfo  {journal} {Phys. Rev. B}\ }\textbf {\bibinfo {volume}
  {66}},\ \bibinfo {pages} {144501} (\bibinfo {year} {2002})}\BibitemShut
  {NoStop}%
\bibitem [{\citenamefont {Di~Pietro}\ \emph {et~al.}(2016)\citenamefont
  {Di~Pietro}, \citenamefont {Komargodski}, \citenamefont {Shamir},\ and\
  \citenamefont {Stamou}}]{DiPietro2015qed}%
  \BibitemOpen
  \bibfield  {author} {\bibinfo {author} {\bibfnamefont {L.}~\bibnamefont
  {Di~Pietro}}, \bibinfo {author} {\bibfnamefont {Z.}~\bibnamefont
  {Komargodski}}, \bibinfo {author} {\bibfnamefont {I.}~\bibnamefont
  {Shamir}},\ and\ \bibinfo {author} {\bibfnamefont {E.}~\bibnamefont
  {Stamou}},\ }\bibfield  {title} {\bibinfo {title} {Quantum electrodynamics in
  d=3 from the $\varepsilon$ expansion},\ }\href
  {https://journals.aps.org/prl/abstract/10.1103/PhysRevLett.116.131601}
  {\bibfield  {journal} {\bibinfo  {journal} {Phys. Rev. Lett.}\ }\textbf
  {\bibinfo {volume} {116}},\ \bibinfo {pages} {131601} (\bibinfo {year}
  {2016})}\BibitemShut {NoStop}%
\bibitem [{\citenamefont {Giombi}\ \emph
  {et~al.}(2016{\natexlab{a}})\citenamefont {Giombi}, \citenamefont
  {Tarnopolsky},\ and\ \citenamefont {Klebanov}}]{Giombi2016currentsQED}%
  \BibitemOpen
  \bibfield  {author} {\bibinfo {author} {\bibfnamefont {S.}~\bibnamefont
  {Giombi}}, \bibinfo {author} {\bibfnamefont {G.}~\bibnamefont
  {Tarnopolsky}},\ and\ \bibinfo {author} {\bibfnamefont {I.~R.}\ \bibnamefont
  {Klebanov}},\ }\bibfield  {title} {\bibinfo {title} {On {$C_J$} and {$C_T$}
  in conformal {QED}},\ }\href {https://doi.org/10.1007/JHEP08(2016)156}
  {\bibfield  {journal} {\bibinfo  {journal} {J. High Energy Phys.}\ }\textbf
  {\bibinfo {volume} {2016}}\bibfield  {number} {\bibinfo  {number} { (08)},\
  \bibinfo {pages} {156}},\ }\Eprint {https://arxiv.org/abs/1602.01076}
  {arXiv:1602.01076 [hep-th]} \BibitemShut {NoStop}%
\bibitem [{\citenamefont {Di~Pietro}\ and\ \citenamefont
  {Stamou}(2017)}]{DiPietro2017scalingQED3}%
  \BibitemOpen
  \bibfield  {author} {\bibinfo {author} {\bibfnamefont {L.}~\bibnamefont
  {Di~Pietro}}\ and\ \bibinfo {author} {\bibfnamefont {E.}~\bibnamefont
  {Stamou}},\ }\bibfield  {title} {\bibinfo {title} {Scaling dimensions in
  {QED}$_3$ from the {$\epsilon$}-expansion},\ }\href
  {https://doi.org/10.1007/JHEP12(2017)054} {\bibfield  {journal} {\bibinfo
  {journal} {J. High Energy Phys.}\ }\textbf {\bibinfo {volume}
  {2017}}\bibfield  {number} {\bibinfo  {number} { (12)},\ \bibinfo {pages}
  {054}},\ }\Eprint {https://arxiv.org/abs/1708.03740} {arXiv:1708.03740
  [hep-th]} \BibitemShut {NoStop}%
\bibitem [{\citenamefont {Hermele}\ \emph {et~al.}(2004)\citenamefont
  {Hermele}, \citenamefont {Senthil}, \citenamefont {Fisher}, \citenamefont
  {Lee}, \citenamefont {Nagaosa},\ and\ \citenamefont
  {Wen}}]{Hermele2004stability}%
  \BibitemOpen
  \bibfield  {author} {\bibinfo {author} {\bibfnamefont {M.}~\bibnamefont
  {Hermele}}, \bibinfo {author} {\bibfnamefont {T.}~\bibnamefont {Senthil}},
  \bibinfo {author} {\bibfnamefont {M.~P.~A.}\ \bibnamefont {Fisher}}, \bibinfo
  {author} {\bibfnamefont {P.~A.}\ \bibnamefont {Lee}}, \bibinfo {author}
  {\bibfnamefont {N.}~\bibnamefont {Nagaosa}},\ and\ \bibinfo {author}
  {\bibfnamefont {X.-G.}\ \bibnamefont {Wen}},\ }\bibfield  {title} {\bibinfo
  {title} {Stability of {$U(1)$} spin liquids in two dimensions},\ }\href
  {https://doi.org/10.1103/PhysRevB.70.214437} {\bibfield  {journal} {\bibinfo
  {journal} {Phys. Rev. B}\ }\textbf {\bibinfo {volume} {70}},\ \bibinfo
  {pages} {214437} (\bibinfo {year} {2004})}\BibitemShut {NoStop}%
\bibitem [{\citenamefont {Ran}\ \emph {et~al.}(2007)\citenamefont {Ran},
  \citenamefont {Hermele}, \citenamefont {Lee},\ and\ \citenamefont
  {Wen}}]{Ran2007kagomeDSL}%
  \BibitemOpen
  \bibfield  {author} {\bibinfo {author} {\bibfnamefont {Y.}~\bibnamefont
  {Ran}}, \bibinfo {author} {\bibfnamefont {M.}~\bibnamefont {Hermele}},
  \bibinfo {author} {\bibfnamefont {P.~A.}\ \bibnamefont {Lee}},\ and\ \bibinfo
  {author} {\bibfnamefont {X.-G.}\ \bibnamefont {Wen}},\ }\bibfield  {title}
  {\bibinfo {title} {Projected-wave-function study of the spin-1/2 heisenberg
  model on the kagom\'e lattice},\ }\href
  {https://doi.org/10.1103/PhysRevLett.98.117205} {\bibfield  {journal}
  {\bibinfo  {journal} {Phys. Rev. Lett.}\ }\textbf {\bibinfo {volume} {98}},\
  \bibinfo {pages} {117205} (\bibinfo {year} {2007})}\BibitemShut {NoStop}%
\bibitem [{\citenamefont {Hermele}\ \emph {et~al.}(2008)\citenamefont
  {Hermele}, \citenamefont {Ran}, \citenamefont {Lee},\ and\ \citenamefont
  {Wen}}]{Hermele2008kagomeASL}%
  \BibitemOpen
  \bibfield  {author} {\bibinfo {author} {\bibfnamefont {M.}~\bibnamefont
  {Hermele}}, \bibinfo {author} {\bibfnamefont {Y.}~\bibnamefont {Ran}},
  \bibinfo {author} {\bibfnamefont {P.~A.}\ \bibnamefont {Lee}},\ and\ \bibinfo
  {author} {\bibfnamefont {X.-G.}\ \bibnamefont {Wen}},\ }\bibfield  {title}
  {\bibinfo {title} {Properties of an algebraic spin liquid on the kagome
  lattice},\ }\href {https://doi.org/10.1103/PhysRevB.77.224413} {\bibfield
  {journal} {\bibinfo  {journal} {Phys. Rev. B}\ }\textbf {\bibinfo {volume}
  {77}},\ \bibinfo {pages} {224413} (\bibinfo {year} {2008})}\BibitemShut
  {NoStop}%
\bibitem [{\citenamefont {He}\ \emph {et~al.}(2017)\citenamefont {He},
  \citenamefont {Zaletel}, \citenamefont {Oshikawa},\ and\ \citenamefont
  {Pollmann}}]{He2017kagomeDSL}%
  \BibitemOpen
  \bibfield  {author} {\bibinfo {author} {\bibfnamefont {Y.-C.}\ \bibnamefont
  {He}}, \bibinfo {author} {\bibfnamefont {M.~P.}\ \bibnamefont {Zaletel}},
  \bibinfo {author} {\bibfnamefont {M.}~\bibnamefont {Oshikawa}},\ and\
  \bibinfo {author} {\bibfnamefont {F.}~\bibnamefont {Pollmann}},\ }\bibfield
  {title} {\bibinfo {title} {Signatures of dirac cones in a {DMRG} study of the
  kagome heisenberg model},\ }\href {https://doi.org/10.1103/PhysRevX.7.031020}
  {\bibfield  {journal} {\bibinfo  {journal} {Phys. Rev. X}\ }\textbf {\bibinfo
  {volume} {7}},\ \bibinfo {pages} {031020} (\bibinfo {year}
  {2017})}\BibitemShut {NoStop}%
\bibitem [{\citenamefont {Hu}\ \emph {et~al.}(2019)\citenamefont {Hu},
  \citenamefont {Zhu}, \citenamefont {Eggert},\ and\ \citenamefont
  {He}}]{Hu2019triangularDSL}%
  \BibitemOpen
  \bibfield  {author} {\bibinfo {author} {\bibfnamefont {S.}~\bibnamefont
  {Hu}}, \bibinfo {author} {\bibfnamefont {W.}~\bibnamefont {Zhu}}, \bibinfo
  {author} {\bibfnamefont {S.}~\bibnamefont {Eggert}},\ and\ \bibinfo {author}
  {\bibfnamefont {Y.-C.}\ \bibnamefont {He}},\ }\bibfield  {title} {\bibinfo
  {title} {Dirac spin liquid on the spin-1/2 triangular heisenberg
  antiferromagnet},\ }\href {https://doi.org/10.1103/PhysRevLett.123.207203}
  {\bibfield  {journal} {\bibinfo  {journal} {Phys. Rev. Lett.}\ }\textbf
  {\bibinfo {volume} {123}},\ \bibinfo {pages} {207203} (\bibinfo {year}
  {2019})}\BibitemShut {NoStop}%
\bibitem [{\citenamefont {Lee}\ \emph {et~al.}(2006)\citenamefont {Lee},
  \citenamefont {Nagaosa},\ and\ \citenamefont {Wen}}]{lee2006doping}%
  \BibitemOpen
  \bibfield  {author} {\bibinfo {author} {\bibfnamefont {P.~A.}\ \bibnamefont
  {Lee}}, \bibinfo {author} {\bibfnamefont {N.}~\bibnamefont {Nagaosa}},\ and\
  \bibinfo {author} {\bibfnamefont {X.-G.}\ \bibnamefont {Wen}},\ }\bibfield
  {title} {\bibinfo {title} {Doping a mott insulator: Physics of
  high-temperature superconductivity},\ }\href
  {https://journals.aps.org/rmp/abstract/10.1103/RevModPhys.78.17} {\bibfield
  {journal} {\bibinfo  {journal} {Reviews of modern physics}\ }\textbf
  {\bibinfo {volume} {78}},\ \bibinfo {pages} {17} (\bibinfo {year}
  {2006})}\BibitemShut {NoStop}%
\bibitem [{\citenamefont {Senthil}\ \emph {et~al.}(2004)\citenamefont
  {Senthil}, \citenamefont {Vishwanath}, \citenamefont {Balents}, \citenamefont
  {Sachdev},\ and\ \citenamefont {Fisher}}]{senthil2004deconfined}%
  \BibitemOpen
  \bibfield  {author} {\bibinfo {author} {\bibfnamefont {T.}~\bibnamefont
  {Senthil}}, \bibinfo {author} {\bibfnamefont {A.}~\bibnamefont {Vishwanath}},
  \bibinfo {author} {\bibfnamefont {L.}~\bibnamefont {Balents}}, \bibinfo
  {author} {\bibfnamefont {S.}~\bibnamefont {Sachdev}},\ and\ \bibinfo {author}
  {\bibfnamefont {M.~P.}\ \bibnamefont {Fisher}},\ }\bibfield  {title}
  {\bibinfo {title} {Deconfined quantum critical points},\ }\href
  {https://www.science.org/doi/abs/10.1126/science.1091806} {\bibfield
  {journal} {\bibinfo  {journal} {Science}\ }\textbf {\bibinfo {volume}
  {303}},\ \bibinfo {pages} {1490} (\bibinfo {year} {2004})}\BibitemShut
  {NoStop}%
\bibitem [{\citenamefont {Wang}\ \emph {et~al.}(2017)\citenamefont {Wang},
  \citenamefont {Nahum}, \citenamefont {Metlitski}, \citenamefont {Xu},\ and\
  \citenamefont {Senthil}}]{wang2017deconfined}%
  \BibitemOpen
  \bibfield  {author} {\bibinfo {author} {\bibfnamefont {C.}~\bibnamefont
  {Wang}}, \bibinfo {author} {\bibfnamefont {A.}~\bibnamefont {Nahum}},
  \bibinfo {author} {\bibfnamefont {M.~A.}\ \bibnamefont {Metlitski}}, \bibinfo
  {author} {\bibfnamefont {C.}~\bibnamefont {Xu}},\ and\ \bibinfo {author}
  {\bibfnamefont {T.}~\bibnamefont {Senthil}},\ }\bibfield  {title} {\bibinfo
  {title} {Deconfined quantum critical points: symmetries and dualities},\
  }\href {https://journals.aps.org/prx/abstract/10.1103/PhysRevX.7.031051}
  {\bibfield  {journal} {\bibinfo  {journal} {Physical Review X}\ }\textbf
  {\bibinfo {volume} {7}},\ \bibinfo {pages} {031051} (\bibinfo {year}
  {2017})}\BibitemShut {NoStop}%
\bibitem [{\citenamefont {Gingras}\ and\ \citenamefont
  {McClarty}(2014)}]{gingras2014quantum}%
  \BibitemOpen
  \bibfield  {author} {\bibinfo {author} {\bibfnamefont {M.~J.}\ \bibnamefont
  {Gingras}}\ and\ \bibinfo {author} {\bibfnamefont {P.~A.}\ \bibnamefont
  {McClarty}},\ }\bibfield  {title} {\bibinfo {title} {Quantum spin ice: a
  search for gapless quantum spin liquids in pyrochlore magnets},\ }\href
  {https://iopscience.iop.org/article/10.1088/0034-4885/77/5/056501/}
  {\bibfield  {journal} {\bibinfo  {journal} {Reports on Progress in Physics}\
  }\textbf {\bibinfo {volume} {77}},\ \bibinfo {pages} {056501} (\bibinfo
  {year} {2014})}\BibitemShut {NoStop}%
\bibitem [{\citenamefont {Wan}\ \emph {et~al.}(2026)\citenamefont {Wan},
  \citenamefont {Jiang}, \citenamefont {Zou}, \citenamefont {Zhang},\ and\
  \citenamefont {Jian}}]{wan2026quantum}%
  \BibitemOpen
  \bibfield  {author} {\bibinfo {author} {\bibfnamefont {Z.-Q.}\ \bibnamefont
  {Wan}}, \bibinfo {author} {\bibfnamefont {H.}~\bibnamefont {Jiang}}, \bibinfo
  {author} {\bibfnamefont {X.}~\bibnamefont {Zou}}, \bibinfo {author}
  {\bibfnamefont {S.}~\bibnamefont {Zhang}},\ and\ \bibinfo {author}
  {\bibfnamefont {S.-K.}\ \bibnamefont {Jian}},\ }\bibfield  {title} {\bibinfo
  {title} {Quantum charge-4e superconductivity and deconfined pseudocriticality
  in the attractive su (4) hubbard model},\ }\href@noop {} {\bibfield
  {journal} {\bibinfo  {journal} {arXiv preprint arXiv:2604.14289}\ } (\bibinfo
  {year} {2026})}\BibitemShut {NoStop}%
\bibitem [{\citenamefont {Teber}(2012)}]{Teber2012reducedQED}%
  \BibitemOpen
  \bibfield  {author} {\bibinfo {author} {\bibfnamefont {S.}~\bibnamefont
  {Teber}},\ }\bibfield  {title} {\bibinfo {title} {Electromagnetic current
  correlations in reduced quantum electrodynamics},\ }\href
  {https://doi.org/10.1103/PhysRevD.86.025005} {\bibfield  {journal} {\bibinfo
  {journal} {Phys. Rev. D}\ }\textbf {\bibinfo {volume} {86}},\ \bibinfo
  {pages} {025005} (\bibinfo {year} {2012})}\BibitemShut {NoStop}%
\bibitem [{\citenamefont {Hsiao}\ and\ \citenamefont
  {Son}(2017)}]{Hsiao2017mixedQED}%
  \BibitemOpen
  \bibfield  {author} {\bibinfo {author} {\bibfnamefont {W.-H.}\ \bibnamefont
  {Hsiao}}\ and\ \bibinfo {author} {\bibfnamefont {D.~T.}\ \bibnamefont
  {Son}},\ }\bibfield  {title} {\bibinfo {title} {Duality and universal
  transport in mixed-dimension electrodynamics},\ }\href
  {https://doi.org/10.1103/PhysRevB.96.075127} {\bibfield  {journal} {\bibinfo
  {journal} {Phys. Rev. B}\ }\textbf {\bibinfo {volume} {96}},\ \bibinfo
  {pages} {075127} (\bibinfo {year} {2017})}\BibitemShut {NoStop}%
\bibitem [{\citenamefont {Fraser-Taliente}\ \emph {et~al.}(2025)\citenamefont
  {Fraser-Taliente}, \citenamefont {Herzog},\ and\ \citenamefont
  {Shrestha}}]{FraserTaliente2025nonlocalSchwinger}%
  \BibitemOpen
  \bibfield  {author} {\bibinfo {author} {\bibfnamefont {L.}~\bibnamefont
  {Fraser-Taliente}}, \bibinfo {author} {\bibfnamefont {C.~P.}\ \bibnamefont
  {Herzog}},\ and\ \bibinfo {author} {\bibfnamefont {A.}~\bibnamefont
  {Shrestha}},\ }\bibfield  {title} {\bibinfo {title} {A nonlocal schwinger
  model},\ }\href {https://doi.org/10.1007/JHEP06(2025)252} {\bibfield
  {journal} {\bibinfo  {journal} {J. High Energy Phys.}\ }\textbf {\bibinfo
  {volume} {2025}}\bibfield  {number} {\bibinfo  {number} { (06)},\ \bibinfo
  {pages} {252}},\ }\Eprint {https://arxiv.org/abs/2412.02514}
  {arXiv:2412.02514 [hep-th]} \BibitemShut {NoStop}%
\bibitem [{\citenamefont {Herzog}\ and\ \citenamefont
  {Huang}(2017)}]{HerzogHuang2017boundaryCentral}%
  \BibitemOpen
  \bibfield  {author} {\bibinfo {author} {\bibfnamefont {C.~P.}\ \bibnamefont
  {Herzog}}\ and\ \bibinfo {author} {\bibfnamefont {K.-W.}\ \bibnamefont
  {Huang}},\ }\bibfield  {title} {\bibinfo {title} {Boundary conformal field
  theory and a boundary central charge},\ }\href
  {https://doi.org/10.1007/JHEP10(2017)189} {\bibfield  {journal} {\bibinfo
  {journal} {J. High Energy Phys.}\ }\textbf {\bibinfo {volume}
  {2017}}\bibfield  {number} {\bibinfo  {number} { (10)},\ \bibinfo {pages}
  {189}},\ }\Eprint {https://arxiv.org/abs/1707.06224} {arXiv:1707.06224
  [hep-th]} \BibitemShut {NoStop}%
\bibitem [{\citenamefont {Herzog}\ \emph {et~al.}(2018)\citenamefont {Herzog},
  \citenamefont {Huang}, \citenamefont {Shamir},\ and\ \citenamefont
  {Virrueta}}]{Herzog2018grapheneBCFT}%
  \BibitemOpen
  \bibfield  {author} {\bibinfo {author} {\bibfnamefont {C.~P.}\ \bibnamefont
  {Herzog}}, \bibinfo {author} {\bibfnamefont {K.-W.}\ \bibnamefont {Huang}},
  \bibinfo {author} {\bibfnamefont {I.}~\bibnamefont {Shamir}},\ and\ \bibinfo
  {author} {\bibfnamefont {J.}~\bibnamefont {Virrueta}},\ }\bibfield  {title}
  {\bibinfo {title} {Superconformal models for graphene and boundary central
  charges},\ }\href {https://doi.org/10.1007/JHEP09(2018)161} {\bibfield
  {journal} {\bibinfo  {journal} {JHEP}\ }\textbf {\bibinfo {volume} {09}},\
  \bibinfo {pages} {161}},\ \Eprint {https://arxiv.org/abs/1807.01700}
  {arXiv:1807.01700 [hep-th]} \BibitemShut {NoStop}%
\bibitem [{\citenamefont {Di~Pietro}\ \emph {et~al.}(2019)\citenamefont
  {Di~Pietro}, \citenamefont {Gaiotto}, \citenamefont {Lauria},\ and\
  \citenamefont {Wu}}]{DiPietro2019bdygauge}%
  \BibitemOpen
  \bibfield  {author} {\bibinfo {author} {\bibfnamefont {L.}~\bibnamefont
  {Di~Pietro}}, \bibinfo {author} {\bibfnamefont {D.}~\bibnamefont {Gaiotto}},
  \bibinfo {author} {\bibfnamefont {E.}~\bibnamefont {Lauria}},\ and\ \bibinfo
  {author} {\bibfnamefont {J.}~\bibnamefont {Wu}},\ }\bibfield  {title}
  {\bibinfo {title} {3d abelian gauge theories at the boundary},\ }\href
  {https://link.springer.com/article/10.1007/JHEP05(2019)091} {\bibfield
  {journal} {\bibinfo  {journal} {JHEP}\ }\textbf {\bibinfo {volume} {05}},\
  \bibinfo {pages} {091}}\BibitemShut {NoStop}%
\bibitem [{\citenamefont {Bartlett-Tisdall}\ \emph {et~al.}(2024)\citenamefont
  {Bartlett-Tisdall}, \citenamefont {Herzog},\ and\ \citenamefont
  {Schaub}}]{BartlettTisdall2024boundaryQED}%
  \BibitemOpen
  \bibfield  {author} {\bibinfo {author} {\bibfnamefont {S.}~\bibnamefont
  {Bartlett-Tisdall}}, \bibinfo {author} {\bibfnamefont {C.~P.}\ \bibnamefont
  {Herzog}},\ and\ \bibinfo {author} {\bibfnamefont {V.}~\bibnamefont
  {Schaub}},\ }\bibfield  {title} {\bibinfo {title} {Bootstrapping boundary
  {QED}. part {I}},\ }\href {https://doi.org/10.1007/JHEP05(2024)235}
  {\bibfield  {journal} {\bibinfo  {journal} {J. High Energy Phys.}\ }\textbf
  {\bibinfo {volume} {2024}}\bibfield  {number} {\bibinfo  {number} { (05)},\
  \bibinfo {pages} {235}},\ }\Eprint {https://arxiv.org/abs/2312.07692}
  {arXiv:2312.07692 [hep-th]} \BibitemShut {NoStop}%
\bibitem [{\citenamefont {Xu}\ \emph {et~al.}(2019)\citenamefont {Xu},
  \citenamefont {Qi}, \citenamefont {Zhang}, \citenamefont {Assaad},
  \citenamefont {Xu},\ and\ \citenamefont {Meng}}]{Xu2018qmc}%
  \BibitemOpen
  \bibfield  {author} {\bibinfo {author} {\bibfnamefont {X.~Y.}\ \bibnamefont
  {Xu}}, \bibinfo {author} {\bibfnamefont {Y.}~\bibnamefont {Qi}}, \bibinfo
  {author} {\bibfnamefont {L.}~\bibnamefont {Zhang}}, \bibinfo {author}
  {\bibfnamefont {F.~F.}\ \bibnamefont {Assaad}}, \bibinfo {author}
  {\bibfnamefont {C.}~\bibnamefont {Xu}},\ and\ \bibinfo {author}
  {\bibfnamefont {Z.~Y.}\ \bibnamefont {Meng}},\ }\bibfield  {title} {\bibinfo
  {title} {Monte carlo study of lattice compact quantum electrodynamics with
  fermionic matter: The parent state of quantum phases},\ }\href
  {https://doi.org/10.1103/PhysRevX.9.021022} {\bibfield  {journal} {\bibinfo
  {journal} {Phys. Rev. X}\ }\textbf {\bibinfo {volume} {9}},\ \bibinfo {pages}
  {021022} (\bibinfo {year} {2019})}\BibitemShut {NoStop}%
\bibitem [{\citenamefont {Feng}\ \emph {et~al.}(2026)\citenamefont {Feng},
  \citenamefont {Chen},\ and\ \citenamefont {Meng}}]{Feng2026scalable}%
  \BibitemOpen
  \bibfield  {author} {\bibinfo {author} {\bibfnamefont {K.}~\bibnamefont
  {Feng}}, \bibinfo {author} {\bibfnamefont {C.}~\bibnamefont {Chen}},\ and\
  \bibinfo {author} {\bibfnamefont {Z.~Y.}\ \bibnamefont {Meng}},\ }\bibfield
  {title} {\bibinfo {title} {Scalable hybrid quantum monte carlo simulation of
  {$U(1)$} gauge field coupled to fermions on {GPU}},\ }\href
  {https://doi.org/10.21468/SciPostPhys.20.2.060} {\bibfield  {journal}
  {\bibinfo  {journal} {SciPost Phys.}\ }\textbf {\bibinfo {volume} {20}},\
  \bibinfo {pages} {060} (\bibinfo {year} {2026})},\ \Eprint
  {https://arxiv.org/abs/2508.16298} {arXiv:2508.16298 [cond-mat.str-el]}
  \BibitemShut {NoStop}%
\bibitem [{\citenamefont {Celi}\ \emph {et~al.}(2020)\citenamefont {Celi},
  \citenamefont {Vermersch}, \citenamefont {Viyuela}, \citenamefont {Pichler},
  \citenamefont {Lukin},\ and\ \citenamefont {Zoller}}]{Celi2019lqy}%
  \BibitemOpen
  \bibfield  {author} {\bibinfo {author} {\bibfnamefont {A.}~\bibnamefont
  {Celi}}, \bibinfo {author} {\bibfnamefont {B.}~\bibnamefont {Vermersch}},
  \bibinfo {author} {\bibfnamefont {O.}~\bibnamefont {Viyuela}}, \bibinfo
  {author} {\bibfnamefont {H.}~\bibnamefont {Pichler}}, \bibinfo {author}
  {\bibfnamefont {M.~D.}\ \bibnamefont {Lukin}},\ and\ \bibinfo {author}
  {\bibfnamefont {P.}~\bibnamefont {Zoller}},\ }\bibfield  {title} {\bibinfo
  {title} {Emerging two-dimensional gauge theories in rydberg configurable
  arrays},\ }\href
  {https://journals.aps.org/prx/abstract/10.1103/PhysRevX.10.021057} {\bibfield
   {journal} {\bibinfo  {journal} {Phys. Rev. X}\ }\textbf {\bibinfo {volume}
  {10}},\ \bibinfo {pages} {021057} (\bibinfo {year} {2020})}\BibitemShut
  {NoStop}%
\bibitem [{\citenamefont {Giombi}\ \emph
  {et~al.}(2016{\natexlab{b}})\citenamefont {Giombi}, \citenamefont
  {Klebanov},\ and\ \citenamefont {Tarnopolsky}}]{Giombi2016conformalQED}%
  \BibitemOpen
  \bibfield  {author} {\bibinfo {author} {\bibfnamefont {S.}~\bibnamefont
  {Giombi}}, \bibinfo {author} {\bibfnamefont {I.~R.}\ \bibnamefont
  {Klebanov}},\ and\ \bibinfo {author} {\bibfnamefont {G.}~\bibnamefont
  {Tarnopolsky}},\ }\bibfield  {title} {\bibinfo {title} {Conformal {QED}$_d$,
  {$F$}-theorem and the {$\epsilon$} expansion},\ }\href
  {https://doi.org/10.1088/1751-8113/49/13/135403} {\bibfield  {journal}
  {\bibinfo  {journal} {J. Phys. A: Math. Theor.}\ }\textbf {\bibinfo {volume}
  {49}},\ \bibinfo {pages} {135403} (\bibinfo {year} {2016}{\natexlab{b}})},\
  \Eprint {https://arxiv.org/abs/1508.06354} {arXiv:1508.06354 [hep-th]}
  \BibitemShut {NoStop}%
\bibitem [{\citenamefont {Thorngren}\ \emph {et~al.}(2023)\citenamefont
  {Thorngren}, \citenamefont {Rakovszky}, \citenamefont {Verresen},\ and\
  \citenamefont {Vishwanath}}]{thorngren2023higgs}%
  \BibitemOpen
  \bibfield  {author} {\bibinfo {author} {\bibfnamefont {R.}~\bibnamefont
  {Thorngren}}, \bibinfo {author} {\bibfnamefont {T.}~\bibnamefont
  {Rakovszky}}, \bibinfo {author} {\bibfnamefont {R.}~\bibnamefont
  {Verresen}},\ and\ \bibinfo {author} {\bibfnamefont {A.}~\bibnamefont
  {Vishwanath}},\ }\bibfield  {title} {\bibinfo {title} {Higgs condensates are
  symmetry-protected topological phases: Ii. $u(1)$ gauge theory and
  superconductors},\ }\href {https://arxiv.org/abs/2303.08136} {\bibfield
  {journal} {\bibinfo  {journal} {arXiv:2303.08136}\ } (\bibinfo {year}
  {2023})}\BibitemShut {NoStop}%
\bibitem [{\citenamefont {Verresen}\ \emph {et~al.}(2024)\citenamefont
  {Verresen}, \citenamefont {Borla}, \citenamefont {Vishwanath}, \citenamefont
  {Moroz},\ and\ \citenamefont {Thorngren}}]{verresen2024higgs}%
  \BibitemOpen
  \bibfield  {author} {\bibinfo {author} {\bibfnamefont {R.}~\bibnamefont
  {Verresen}}, \bibinfo {author} {\bibfnamefont {U.}~\bibnamefont {Borla}},
  \bibinfo {author} {\bibfnamefont {A.}~\bibnamefont {Vishwanath}}, \bibinfo
  {author} {\bibfnamefont {S.}~\bibnamefont {Moroz}},\ and\ \bibinfo {author}
  {\bibfnamefont {R.}~\bibnamefont {Thorngren}},\ }\bibfield  {title} {\bibinfo
  {title} {Higgs condensates are symmetry-protected topological phases: I.
  discrete symmetries},\ }\href {https://arxiv.org/abs/2211.01376} {\bibfield
  {journal} {\bibinfo  {journal} {arXiv:2211.01376}\ } (\bibinfo {year}
  {2024})}\BibitemShut {NoStop}%
\bibitem [{\citenamefont {Chung}\ \emph {et~al.}(2025)\citenamefont {Chung},
  \citenamefont {Flores-Calderón}, \citenamefont {Torres}, \citenamefont
  {Ribeiro}, \citenamefont {Moroz},\ and\ \citenamefont
  {McClarty}}]{chung2024higgs}%
  \BibitemOpen
  \bibfield  {author} {\bibinfo {author} {\bibfnamefont {K.~T.~K.}\
  \bibnamefont {Chung}}, \bibinfo {author} {\bibfnamefont {R.}~\bibnamefont
  {Flores-Calderón}}, \bibinfo {author} {\bibfnamefont {R.~C.}\ \bibnamefont
  {Torres}}, \bibinfo {author} {\bibfnamefont {P.}~\bibnamefont {Ribeiro}},
  \bibinfo {author} {\bibfnamefont {S.}~\bibnamefont {Moroz}},\ and\ \bibinfo
  {author} {\bibfnamefont {P.}~\bibnamefont {McClarty}},\ }\bibfield  {title}
  {\bibinfo {title} {Higgs phases and boundary criticality},\ }\href
  {https://doi.org/10.21468/SciPostPhys.19.4.105} {\bibfield  {journal}
  {\bibinfo  {journal} {SciPost Phys.}\ }\textbf {\bibinfo {volume} {19}},\
  \bibinfo {pages} {105} (\bibinfo {year} {2025})}\BibitemShut {NoStop}%
\bibitem [{\citenamefont {Nishioka}\ \emph {et~al.}(2022)\citenamefont
  {Nishioka}, \citenamefont {Okuyama},\ and\ \citenamefont
  {Shimamori}}]{nishioka2022method}%
  \BibitemOpen
  \bibfield  {author} {\bibinfo {author} {\bibfnamefont {T.}~\bibnamefont
  {Nishioka}}, \bibinfo {author} {\bibfnamefont {Y.}~\bibnamefont {Okuyama}},\
  and\ \bibinfo {author} {\bibfnamefont {S.}~\bibnamefont {Shimamori}},\
  }\bibfield  {title} {\bibinfo {title} {Method of images in defect conformal
  field theories},\ }\href {https://doi.org/10.1103/PhysRevD.106.L081701}
  {\bibfield  {journal} {\bibinfo  {journal} {Phys. Rev. D}\ }\textbf {\bibinfo
  {volume} {106}},\ \bibinfo {pages} {L081701} (\bibinfo {year}
  {2022})}\BibitemShut {NoStop}%
\bibitem [{\citenamefont {Feldmeier}\ \emph {et~al.}(2020)\citenamefont
  {Feldmeier}, \citenamefont {Natori}, \citenamefont {Knap},\ and\
  \citenamefont {Knolle}}]{feldmeier2020local}%
  \BibitemOpen
  \bibfield  {author} {\bibinfo {author} {\bibfnamefont {J.}~\bibnamefont
  {Feldmeier}}, \bibinfo {author} {\bibfnamefont {W.}~\bibnamefont {Natori}},
  \bibinfo {author} {\bibfnamefont {M.}~\bibnamefont {Knap}},\ and\ \bibinfo
  {author} {\bibfnamefont {J.}~\bibnamefont {Knolle}},\ }\bibfield  {title}
  {\bibinfo {title} {Local probes for charge-neutral edge states in
  two-dimensional quantum magnets},\ }\href
  {https://doi.org/10.1103/PhysRevB.102.134423} {\bibfield  {journal} {\bibinfo
   {journal} {Phys. Rev. B}\ }\textbf {\bibinfo {volume} {102}},\ \bibinfo
  {pages} {134423} (\bibinfo {year} {2020})}\BibitemShut {NoStop}%
\bibitem [{\citenamefont {K{\"o}nig}\ \emph {et~al.}(2020)\citenamefont
  {K{\"o}nig}, \citenamefont {Randeria},\ and\ \citenamefont
  {J{\"a}ck}}]{konig2020tunneling}%
  \BibitemOpen
  \bibfield  {author} {\bibinfo {author} {\bibfnamefont {E.~J.}\ \bibnamefont
  {K{\"o}nig}}, \bibinfo {author} {\bibfnamefont {M.~T.}\ \bibnamefont
  {Randeria}},\ and\ \bibinfo {author} {\bibfnamefont {B.}~\bibnamefont
  {J{\"a}ck}},\ }\bibfield  {title} {\bibinfo {title} {Tunneling spectroscopy
  of quantum spin liquids},\ }\href
  {https://doi.org/10.1103/PhysRevLett.125.267206} {\bibfield  {journal}
  {\bibinfo  {journal} {Phys. Rev. Lett.}\ }\textbf {\bibinfo {volume} {125}},\
  \bibinfo {pages} {267206} (\bibinfo {year} {2020})}\BibitemShut {NoStop}%
\bibitem [{\citenamefont {Udagawa}\ \emph {et~al.}(2021)\citenamefont
  {Udagawa}, \citenamefont {Takayoshi},\ and\ \citenamefont
  {Oka}}]{udagawa2021scanning}%
  \BibitemOpen
  \bibfield  {author} {\bibinfo {author} {\bibfnamefont {M.}~\bibnamefont
  {Udagawa}}, \bibinfo {author} {\bibfnamefont {S.}~\bibnamefont {Takayoshi}},\
  and\ \bibinfo {author} {\bibfnamefont {T.}~\bibnamefont {Oka}},\ }\bibfield
  {title} {\bibinfo {title} {Scanning tunneling microscopy as a single majorana
  detector of kitaev's chiral spin liquid},\ }\href
  {https://doi.org/10.1103/PhysRevLett.126.127201} {\bibfield  {journal}
  {\bibinfo  {journal} {Phys. Rev. Lett.}\ }\textbf {\bibinfo {volume} {126}},\
  \bibinfo {pages} {127201} (\bibinfo {year} {2021})}\BibitemShut {NoStop}%
\bibitem [{\citenamefont {Zhang}\ \emph {et~al.}(2025)\citenamefont {Zhang},
  \citenamefont {Batista},\ and\ \citenamefont {Hal{\'a}sz}}]{zhang2025edge}%
  \BibitemOpen
  \bibfield  {author} {\bibinfo {author} {\bibfnamefont {S.-S.}\ \bibnamefont
  {Zhang}}, \bibinfo {author} {\bibfnamefont {C.~D.}\ \bibnamefont {Batista}},\
  and\ \bibinfo {author} {\bibfnamefont {G.~B.}\ \bibnamefont {Hal{\'a}sz}},\
  }\bibfield  {title} {\bibinfo {title} {Low-energy edge signatures of the
  kitaev spin liquid},\ }\href {https://doi.org/10.1103/PhysRevB.111.L161104}
  {\bibfield  {journal} {\bibinfo  {journal} {Phys. Rev. B}\ }\textbf {\bibinfo
  {volume} {111}},\ \bibinfo {pages} {L161104} (\bibinfo {year}
  {2025})}\BibitemShut {NoStop}%
\bibitem [{\citenamefont {De~Cesare}\ and\ \citenamefont
  {Giombi}(2026)}]{de2026conformal}%
  \BibitemOpen
  \bibfield  {author} {\bibinfo {author} {\bibfnamefont {F.}~\bibnamefont
  {De~Cesare}}\ and\ \bibinfo {author} {\bibfnamefont {S.}~\bibnamefont
  {Giombi}},\ }\bibfield  {title} {\bibinfo {title} {Conformal qed in ads as a
  bcft},\ }\href {https://arxiv.org/pdf/2607.19464} {\bibfield  {journal}
  {\bibinfo  {journal} {arXiv preprint arXiv:2607.19464}\ } (\bibinfo {year}
  {2026})}\BibitemShut {NoStop}%
\end{thebibliography}%

\onecolumngrid
\setcounter{secnumdepth}{3}
\setcounter{equation}{0}
\setcounter{figure}{0}
\renewcommand{\theequation}{S\arabic{equation}}
\renewcommand{\thefigure}{S\arabic{figure}}
\renewcommand\figurename{Supplementary Figure}
\renewcommand\tablename{Supplementary Table}

\section*{Supplemental Material}

\section{Renormalization group analysis of QED$_3$}
In this section, we review the renormalization group (RG) analysis of massless quantum electrodynamics (QED). In Euclidean spacetime, the action reads
\begin{align}
    S=\int_{\mathcal M} \dd^dx \left[\sum_{j}^N\bar\psi_j\gamma^\mu(\partial_\mu-ieA_\mu)\psi_j+\frac{1}{4}F_{\mu\nu}F_{\mu\nu}\right],
\end{align}
where $F_{\mu\nu}=\partial_\mu A_\nu-\partial_\nu A_\mu$ is the field strength, $\psi_j$ ($j=1,\cdots,N$) are four-component Dirac fermions, and $e$ denotes the gauge coupling.
$\gamma_{\mu}$ are Dirac matrices satisfy $\{\gamma^\mu,\gamma^\nu\}=2\delta^{\mu\nu}$.
We employ dimensional regularization in $d=4-\varepsilon$ spacetime dimensions with minimal subtraction scheme, and work in the Feynman gauge. 
The bare quantities are related to their renormalized ones (subscript $R$) by
\begin{align}
    e=\mu^{\varepsilon/2}Z_1Z_2^{-1}Z_3^{-1/2}e_R,\qquad \psi=\sqrt{Z_2}\psi_R,\qquad A_\mu=\sqrt{Z_3}A_{\mu,R},
\end{align}
where $\mu$ is the renormalization scale, and $Z_1$, $Z_2$, and $Z_3$ denote the vertex, fermion, and gauge-field RG factors, respectively.
At one-loop order, the factors are
\begin{align}
    Z_1=1-\frac{e^2}{8\pi^2\varepsilon}, \qquad Z_2=1-\frac{e^2}{8\pi^2\varepsilon},\qquad Z_3=1-\frac{Ne^2}{6\pi^2\varepsilon}.
\end{align}
The equality $Z_1=Z_2$ is guaranteed by the Ward-Takahashi identity. 
The beta function reads
\begin{align}
    \beta_e\equiv\frac{\dd e}{\dd\log \mu}=-\frac{\varepsilon}{2}e+\frac{Ne^3}{12\pi^2},
\end{align}
which has an infrared-stable fixed point at ${(e^*)}^2=6\pi^2\varepsilon/N$.

\section{Renormalization of boundary QED}
In this section, we formulate QED in the presence of a boundary and derive the corresponding propagators. 
We consider the semi-infinite Euclidean spacetime $\mathcal{M}_+ = \{x_\mu \mid x_{d-1} \ge 0\}$, with the boundary located at $z\equiv x_{d-1} = 0$.
Both the fermion and gauge fields live in $\mathcal M_+$, and the action takes the form,
\begin{align}
S=\int_{\mathcal M_+} \dd^dx \left[\sum_{j}^N\bar\psi_j\gamma^\mu(\partial_\mu-ieA_\mu)\psi_j+\frac{1}{4}F_{\mu\nu}F_{\mu\nu}\right].
\end{align}
Due to the unbroken translational invariance along the $d-1$ directions parallel to the boundary, denoted by $(\tau,\vec{x})$, it is therefore convenient to partial Fourier transform only these coordinates and keep the $z$ in real space.
In this mixed representation, fields depend on the parallel momentum $k=(\omega,\vec{k})$ and on $z$, where $\omega$ is the Euclidean frequency and $\vec{k}=(k_1,\cdots,k_{d-2})$ denotes the spatial momentum vectors parallel to the boundary. 
The action is invariant under the local $U(1)$ gauge transformation,
\begin{align}
\psi(x)\rightarrow e^{ie\alpha(x)} \psi(x),\qquad \bar\psi(x)\to \bar\psi(x)e^{-ie\alpha(x)},\qquad
A_{\mu}(x) \rightarrow A_{\mu}(x) + \partial_{\mu}\alpha(x).
\end{align}
Once boundary conditions are imposed, the allowed behavior of $\alpha(x)$ are constrained, as demonstrated in the main text.

In the presence of the boundary, the Dirac fermion propagator decomposes into bulk and surface contributions,
\begin{align}
G(x, z, z^{\prime}) 
=G^{(b)}(x, z, z^{\prime}) 
+G^{(s)}(x, z, z^{\prime}).
\end{align}
We impose the conformal boundary condition, $-\gamma_z \psi|_{z=0} = \psi|_{z=0}$, which is relevant in condensed matter systems. 
This condition implies the corresponding constraint on the propagator, $-\gamma_z G(x,0,z^{\prime})=G(x,0,z^{\prime})$.
The propagator can then be constructed using the method of images~\cite{nishioka2022method}, 
\begin{align}
    G^{(b)}(k, z,z^{\prime}) 
    =&\Big( \frac{i\slashed{k}}{2 q} 
    -\frac{\gamma_z}{2} \mathrm{sgn}(z-z^{\prime}) \Big) 
    e^{-q|z-z^{\prime}|} \,,  \\
    G^{(s)}(k, z,z^{\prime}) 
    =& -\gamma_z G^{(b)}(k, -z,z^{\prime}) 
    =-\gamma_z \Big( \frac{i\slashed{k}}{2 q} 
    +\frac{\gamma_z}{2} \mathrm{sgn}(z+z^{\prime})  \Big) e^{-q(z+z^{\prime})},
\end{align}
where we introduce the notations $\slashed{k} =\omega \gamma_0 + \vec{k}\cdot \vec{\gamma}$, and $q^2=\omega^2+\vec k^2$.
The bulk propagator $G^{(b)}$ is obtained following the same steps as in Ref.~\cite{jian2025bdygny}.

To acquire the corresponding real-space propagators, we use the partial Fourier transform along the $d-1$ directions parallel to the boundary, which can be evaluated with the general formulas
\begin{align}
    \int {\rm d}^{d-1}x \frac{e^{-i\vec k\cdot\vec x}}{(x^2 + z^2)^\Delta} 
    =& \frac{2^{\frac{d+1}2-\Delta} \pi^{\frac{d-1}2}}{\Gamma(\Delta)} 
    \left( \frac{q}{|z|} \right)^{\Delta - \frac{d-1}2} K_{\Delta - \frac{d-1}2}(q |z|),  \nonumber  \\
    \int {\rm d}^{d-1}x \frac{ x_i e^{-i\vec k\cdot\vec x}}{(x^2 + z^2)^\Delta} 
    =& \frac{2^{\frac{d+1}2-\Delta} \pi^{\frac{d-1}2}}{\Gamma(\Delta)} \left( \frac{q}{|z|} \right)^{\Delta - \frac{d-1}2} 
    \frac{-ik_i |z|}q K_{\Delta - \frac{d+1}2}(q |z|), 
\end{align}
where $x_i$ labels a tangential direction, and $K_\nu$ is the modified Bessel function.
With these formulas, one finds that the mixed representation propagators above correspond to
\begin{align}
G^{(b)}(x, z, z^{\prime}) 
=& \frac{\Gamma(\frac{d}2)}{2\pi^{d/2}} 
\frac{\slashed{x} + (z-z^{\prime}) \gamma_z}{\left(  x^2 + (z-z^{\prime})^2\right)^{d/2}} \,,   \\
G^{(s)}(x, z, z^{\prime}) 
=&-\gamma_z G^{(b)}(x, -z,z^{\prime}) 
=-\frac{\Gamma(\frac{d}2)}{2\pi^{d/2}} \gamma_z \frac{\slashed{x}- (z+z^{\prime}) \gamma_z}{\left( x^2 + (z+z^{\prime})^2\right)^{d/2}} \,,
\end{align}
where we define the notation $\slashed{x}=\tau \gamma_0 +\vec{x}\cdot\vec{\gamma}$, and $x^2=\tau^2+\vec x^2$.

In Feynman gauge, the boundary conditions are implemented by
$\partial_z A_a=A_z=0$ for Neumann boundary conditions and
$A_a=\partial_z A_z=0$ for Dirichlet boundary conditions at $z=0$.
Therefore, the bare gauge field propagator is given by, 
\begin{align}
D_{\mu\nu}(x, z, z^{\prime}) 
=D_{\mu\nu}^{(b)}(x, z, z^{\prime}) 
+w D_{\mu\nu}^{(s)}(x, z, z^{\prime}),
\label{eq:gaugeprop}
\end{align}
with $w$ indicates the Neumann boundary conditions ($w=1$) and Dirichlet boundary condition ($w=-1$).
Here, the bulk and surface parts are written as,
\begin{align}
D_{\mu\nu}^{(b)}(x, z, z^{\prime}) 
=& \frac{\Gamma(\frac{d-2}2)}{4\pi^{d/2}} \frac{ \delta_{\mu\nu}}{\left(x^2+ (z-z^{\prime})^2\right)^{\frac{d-2}2}}  \,,  \\
D_{\mu\nu}^{(s)}(x, z, z^{\prime}) 
=& \frac{\Gamma(\frac{d-2}2)}{4\pi^{d/2}} \frac{R_{\mu\nu}}{\left(x^2 + (z+z^{\prime})^2\right)^{\frac{d-2}2}},
\end{align} 
where $R_{\mu\nu}=\delta_{\mu\nu} -2\delta_{z\mu} \delta_{z\nu}$ implements reflection respect the boundary~\cite{DiPietro2019bdygauge}.
By performing the partial Fourier transformation along the boundary, the bulk and boundary gauge propagators in the mixed representation are,
\begin{align}
D_{\mu\nu}^{(b)}(k,z,z^{\prime}) 
=&\frac{\delta_{\mu\nu}}{2 q} e^{-q|z-z^{\prime}|}  \,, \\
D_{\mu\nu}^{(s)}(k,z,z^{\prime}) 
=&\frac{R_{\mu\nu}}{2 q} e^{-q(z+z^{\prime})} \, .
\end{align}

\subsection{Renormalization of the boundary fermion and gauge field}
We now consider the correction to boundary field $\hat\psi(x)\equiv\psi(x,z=0)$, and $\hat A_\mu(x)\equiv A_\mu(x,z=0)$, where $x$ denotes the coordinates parallel to the boundary.
We begin with the fermion two-point function $\langle \hat\psi(x)\hat{\bar\psi}(x')\rangle$ with both external points restricted to the boundary. 
Its renormalized correlator is defined by $\langle \hat\psi(x)\hat{\bar\psi}(x')\rangle=Z_2Z_{2}^{\partial}\langle\hat\psi_R(x)\hat{\bar{\psi}}_R(x')\rangle$, with the bulk fermion renormalization $\psi=\sqrt Z_2\psi_R$ and boundary fermion renormalization $\hat\psi=\sqrt{Z_2Z_{2}^{\partial}}\hat\psi_R$.
At one-loop order, the correction to the fermion propagator takes the form
\begin{align}
    e^2\int {\rm d}z_1{\rm d}z_2 G(k,0,z_1) \Sigma(k,z_1,z_2) G(k,z_2,0), 
\end{align}
where $k=(\omega,\vec k)$.
The fermionic self-energy is given by
\begin{align}
    \Sigma(k,z_1,z_2)=\sum_{\mu,\nu}\int{\rm d}^{d-1}x e^{-i\vec k\cdot \vec x}\gamma^\mu G(x,z_1,z_2)\gamma^\nu D_{\mu\nu}(-x,z_2,z_1).
\end{align}
The self-energy involves contributions receives contributions of different combinations of bulk (b) and surface (s) parts of the fermion and gauge field propagators, denoted as $G^{(b)}D^{(b)}$, $G^{(s)}D^{(b)}$, $G^{(s)}D^{(s)}$ and $G^{(b)}D^{(s)}$. 
Applying the Feynman parametrization, the integral transforms into
\begin{align}
    G^{(b)} D^{(b)}:\quad&\frac{1}{4\pi^4} \sum_{\mu} 
    {\gamma}_{\mu} \int_0^{1} {\rm d}u 
    \frac{\big(\slashed{x} + (z_1-z_2) \gamma_z\big) \big( u^{\frac{d}{2}-1} (1-u)^{\frac{d}{2}-2} \big)}{(x^2 + |z_1-z_2|^2)^{d-1}}
    {\gamma}_{\mu},  \nonumber  \\
    G^{(s)} D^{(b)}:\quad& -\frac{1}{4\pi^4} \sum_{\mu} 
    {\gamma}_{\mu} \gamma_z \int_0^{1} {\rm d}u \frac{\big(\slashed{x}- (z_1+z_2) \gamma_z\big) \big( u^{\frac{d}{2}-1} (1-u)^{\frac{d}{2}-2} \big)}{(x^2 + \tilde{z}_{+}^2)^{d-1}}
    {\gamma}_{\mu},\nonumber\\
    G^{(s)} D^{(s)}:\quad& -\sum_{\mu\nu} \frac{\Gamma(\frac{d}2)\Gamma(\frac{d-2}2)}{8\pi}
    \gamma_{\mu} 
    \gamma_z \frac{\slashed{x} - (z+z^{\prime}) \gamma_z}{\left(x^2 + (z+z^{\prime})^2 \right)^{d-1}}
    \gamma_{\nu} \big( \delta_{\mu\nu} - 2\delta_{z\mu}\delta_{z\nu}\big) \,,    \nonumber\\
    G^{(b)} D^{(s)}:\quad& \sum_{\mu\nu} \frac{\Gamma(\frac{d}2)\Gamma(\frac{d-2}2)}{8\pi}
    \gamma_{\mu}
    \frac{\slashed{x} + (z-z^{\prime}) \gamma_z}{\left(x^2 + (z-z^{\prime})^2 \right)^{d/2} \left(x^2 + (z+z^{\prime})^2 \right)^{d/2-1}}
    \gamma_{\nu}
    \big( \delta_{\mu\nu} - 2\delta_{z\mu}\delta_{z\nu}\big)  \\
    =& \int_0^1{\rm d}u u^{\frac{d}{2}-1} (1-u)^{\frac{d}{2}-2}
    \sum_{\mu\nu} \frac{\Gamma(\frac{d}2)\Gamma(\frac{d-2}2)}{8\pi}
    \gamma_{\mu}
    \frac{\slashed{x} + (z-z^{\prime}) \gamma_z}{\left(x^2 + \tilde{z}_-^2 \right)^{d-1}}
    \gamma_{\nu}
    \big( \delta_{\mu\nu} - 2\delta_{z\mu}\delta_{z\nu}\big) \,.
\end{align}
The Fourier transform is performed and the divergence of the self-energy reads
\begin{align}
G^{(b)} D^{(b)} &:\qquad \frac{1}{16\pi^2 }
\Big( \frac{i \slashed{k}}{|z_1-z_2|^{1-\varepsilon}}  
-  \gamma_z \frac{z_1-z_2}{|z_1-z_2|^{3-\varepsilon}}\Big),\nonumber    \\
G^{(s)} D^{(b)} &:\qquad \frac{1}{4\pi^2}
\int _{0}^{1} {\rm d}u\,  u^{\frac{d}{2}-1} (1-u)^{\frac{d}{2}-2} 
\frac{z_1+z_2}{|\tilde{z}_+|^{3-\varepsilon}},\nonumber\\
G^{(s)} D^{(s)} &:\qquad \frac{1}{16\pi^2}
\gamma_z\Big( \frac{i \slashed{k}}{|z_1+z_2|^{1-\varepsilon}} 
+\gamma_z \frac{(z_1+z_2)}{|z_1+z_2|^{3-\varepsilon}} \Big), \nonumber\\
G^{(b)} D^{(s)} &:\qquad -\frac{1}{4\pi^2}
\int _{0}^{1} {\rm d}u u^{\frac{d}{2}-1} (1-u)^{\frac{d}{2}-2} \frac{(z_1-z_2)\gamma_z}{|\tilde{z}_-|^{3-\varepsilon}} \,.
\end{align}

Integrating over the normal direction, the contribution from $G^{(b)}D^{(b)}$ reproduces the standard bulk renormalization, $Z_2=1-\frac{e^2}{8\pi^2\varepsilon}$.
Summing all four contributions yields a one-loop ultraviolet divergence proportional to the bare boundary propagator $G(\vec{q}, 0, 0)$.
After subtracting the bulk divergence associated with $Z_2$, the remaining poles are absorbed into the additional boundary counterterm. Defining $Z_2^\partial=1+\delta_2^\partial$
\begin{eqnarray}
\text{Neumann: } \quad\delta_2^{\partial,N} 
= -\frac{e^2}{2\pi^2\varepsilon} - \frac{e^2}{8\pi^2\varepsilon} = -\frac{5e^2}{8\pi^2\varepsilon}\,, \\
\text{Dirichlet:  }\quad\delta_2^{\partial,D} 
= -\frac{e^2}{2\pi^2\varepsilon} + \frac{e^2}{8\pi^2\varepsilon} = -\frac{3e^2}{8\pi^2\varepsilon}\,,
\end{eqnarray}
which imply distinct wavefunction renormalization factor for the boundary fermion field. The resulting RG factors of the boundary fermion are therefore: $Z_{2}^{\partial}=1-5e^2/8\pi^2\varepsilon$ with the Neumann boundary condition, and $Z_2^{\partial}=1-3e^2/8\pi^2\varepsilon$ for the Dirichlet boundary condition, respectively.

We next consider the one-loop correction to the boundary gauge field propagator, governed by the vacuum polarization tensor $\Pi_{\mu\nu}$. The one-loop contribution is calculated from the fermion bubble diagram,
\begin{align}
    D_{\mu\nu}(k,0,0) 
=&\sum_{\nu_1,\nu_2}\int {\rm d}z_1{\rm d}z_2 D_{\mu\nu_1}({k},0,z_1) \Pi_{\nu_1\nu_2}(k,z_1,z_2) 
D_{\nu_2\nu}(k,z_2,0),   \nonumber\\
\Pi_{\nu_1\nu_2}(k,z_1,z_2) 
=&-\int  {\rm d}^{d-1}x e^{-i\vec{k}\cdot\vec{x}} 
\mathrm{Tr}\Big[ \gamma_{\nu_1} G(x,z_1,z_2) 
\gamma_{\nu_2} G(-x,z_2,z_1) \Big].
\end{align}
In the momentum space, the bulk propagator operator takes the following form due to the gauge invariance constraint,
\begin{align}
\Pi_{\mu\nu}(k,k_z) =\int_{-\infty}^{\infty} {{\rm d}z} e^{ik_zz} \Pi_{\mu\nu}(k,0,z) 
=\Pi(k) (k_{\mu}k_{\nu} -k^2),\qquad
\Pi(k) k^2 = \frac{1}{3} \sum_{\mu\nu} \Pi_{\mu\nu}(k,k_z).
\end{align}
In this sense, to employ the gauge symmetry and simplify the calculation, we would like to take summation over the index.
Noting that 
\begin{align}
D_{\mu\nu}(k,z,z^{\prime}) 
=&\frac{1}{4q^2} \int {\rm d}z_1{\rm d}z_2 \Pi_{\mu\nu}(\vec{k},z_1,z_2) 
e^{-q|z-z_1|-q|z_2-z^{\prime}|},
\end{align}
and we can take summation of $\mu,\nu$ to obtain the correction while respect the gauge symmetry.
Decomposing the fermion bubble $\Pi(k,z_1,z_2)$, leads to four contributions $G^{(b)}G^{(b)}$, $G^{(b)}G^{(s)}$, $G^{(s)}G^{(b)}$, and $G^{(s)}G^{(s)}$ for the fermion loop. 
The mixed terms $G^{(b)}G^{(s)}$, $G^{(s)}G^{(b)}$ vanish due to the trace over an odd number of Dirac matrices.
The remaining contributions yield the divergences
\begin{align}
G^{(b)}G^{(b)}:\quad
&\frac{1}{2\pi^2} \Big( \frac{1}{|z_1-z_2|^{3-\varepsilon}} 
-\frac{q^2}{2|z_1-z_2|^{1-\varepsilon}}\Big),  \nonumber\\
G^{(s)}G^{(s)}:\quad
&\frac{1}{2\pi^2}\Big( \frac{1}{|z_1+z_2|^{3-\varepsilon}}
-\frac{q^2}{6|z_1+z_2|^{1-\varepsilon}}\Big).
\end{align}
Evaluated in the bulk regime, the contribution from $G^{(s)}G^{(s)}$ vanishes, while $G^{(b)}G^{(b)}$ reproduces the bulk renormalization $Z_{3}=1-\frac{Ne^2}{6\pi^2\varepsilon}$.
In contrast, when evaluated near the boundary, the surface contribution $G^{(s)}G^{(s)}$ yields leads to, $Z_{3}^{\partial}=1-\frac{Ne^2}{4\pi^2\varepsilon}$, which exhibits specific renormalization compared to the bulk theory.

\subsection{Renormalization of the boundary fermion composite operator}
\begin{figure}
    \centering
    
    \subfigure[]{
    	\includegraphics[width=0.22\columnwidth]{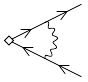}
    	\label{fig:bulk_composite}
    }
    \hspace{0.05\textwidth}
    \subfigure[]{
    	\includegraphics[width=0.22\columnwidth]{bdy_bilinear.pdf}
    	\label{fig:bdy_composite}
    }
    
    \caption{Feynman diagrams relevant to the fermion composite $\bar\psi\gamma^5\psi$. (a) Fermion composite in the bulk theory, $\diamond$ denotes vertex in the bulk. (b) Boundary fermion composite operator, where $\square$ denotes a boundary vertex and the external legs marked by $\circ$ are fixed on the boundary.} 
    \label{fig:fermion_composite}
\end{figure}
In this section, we detail the renormalization of the boundary pseudoscalar bilinear $\hat{\mathcal{O}}_P=\hat{\bar\psi}\gamma^5\hat\psi$. We begin with the corresponding bulk renormalization of this bilinear. 
To account the correction, we introduce the RG factor $\mathcal{O}_P=Z_{\mathcal{O}_P}\mathcal{O}_{P,R}$, where $Z_{\mathcal{O}_P}$ can be decomposited as $Z_{\mathcal{O}_P}=Z_P Z_2$, with $Z_P$ denotes the composite-vertex renormalization. At one-loop order, the $U(1)$ gauge field corrects the vertex via the minimal fermion-gauge interaction, as illustrated in Fig.~\ref{fig:bulk_composite}, leading to
\begin{align}
    (ie)^2\sum_\mu\int\frac{{\rm d}^dp}{(2\pi)^d}\gamma^\mu\frac{i\slashed{p}}{p^2}\gamma^5\frac{i\slashed{p}}{p^2}\gamma^\mu\frac{1}{p^2}=\frac{e^2}{2\pi^2\varepsilon}\gamma^5.
\end{align}
Combining the vertex and wavefunction renormalizations, we obtain $Z_{\mathcal{O}_P}=1+\frac{3e^2}{8\pi^2\varepsilon}$, which leads to the anomalous dimension $\eta_{\mathcal{O}}=\frac{{\rm d}\log Z_\mathcal{O}}{\dd \log\eta}=-\frac{3e^2}{8\pi^2}$.

We next consider the correction to the boundary fermion composite operator. 
We place both the external fermions on the boundary, as shown in Fig.~\ref{fig:bdy_composite}, and the bare three-point function is then $\gamma^5\frac{1+\gamma^3}{2}$, which is independent of the external momentum. 
In the mixed representation, the internal fermion-gauge loop in the one-loop diagram can be decomposed into components parallel and normal to the boundary. The fermionic parts of the one-loop diagram are
\begin{align}
    \sum_{\mu\neq z}G(p,0,z_1)\gamma^\mu G(q,z_1,0)\gamma^5G(q,0,z_2)\gamma^\mu G(p,z_2,0)&=\left(-3\gamma^5(1+\gamma^3)-\gamma^5\frac{(1+\gamma^3)p\cdot q}{pq}\right)e^{-q(z_1+z_2)-p(z_1+z_2)},\nonumber\\
    G(p,0,z_1)\gamma^3 G(q,z_1,0)\gamma^5G(q,0,z_2)\gamma^3 G(p,z_2,0)&=\left(-\gamma^5(1+\gamma^3)+\gamma^5\frac{(1+\gamma^3)p\cdot q}{pq}\right)e^{-q(z_1+z_2)-p(z_1+z_2)}.
\end{align}
Multiplying by the gauge field propagator and taking the limit $p\to0$ leads to the integrations
\begin{align}
    &(ie)^2\gamma^5(1+\gamma^3)\int\frac{r^{2-\varepsilon}\sin\theta\dd r\dd \theta\dd\phi}{(2\pi)^{d-1}}dz_1dz_2\left(\frac{e^{-r|z_1-z_2|}}{2r}+w\frac{e^{-r(z_1+z_2)}}{2r}\right)(-3-\cos\theta)e^{-r(z_1+z_2)}\nonumber\\
    =&\frac{3(2+w)e^2}{8\pi^2\varepsilon}\frac{\gamma^5(1+\gamma^3)}{2},\nonumber\\
    &(ie)^2\gamma^5(1+\gamma^3)\int\frac{r^{2-\varepsilon}\sin\theta\dd r\dd \theta\dd\phi}{(2\pi)^{d-1}}dz_1dz_2\left(\frac{e^{-r|z_1-z_2|}}{2r}-w\frac{e^{-r(z_1+z_2)}}{2r}\right)(-1+\cos\theta)e^{-r(z_1+z_2)}\nonumber\\
    =&\frac{(2-w)e^2}{8\pi^2\varepsilon}\frac{\gamma^5(1+\gamma^3)}{2}.
\end{align}
Here we introduce spherical coordinate for the boundary momentum $(q_0,q_1,q_2)=(r\sin\theta\cos\phi,r\sin\theta\sin\phi,r\cos\theta)$, where $\theta$ is the angle between $p$ and $q$. 
The parameter $w=1$ ($w=-1$) corresponds to Neumann (Dirichlet) boundary condition for the $U(1)$ gauge field. 
Adding together, one can get the RG factor $Z_PZ_{P}^\partial=1+\frac{(4+w)e^2}{4\pi^2\varepsilon}$, with the boundary RG factors defined by $\hat{\mathcal{O}}_P=Z_{\hat{\mathcal O}_P}\hat{\mathcal O}_{P,R}=Z_2Z_2^\partial Z_{P} Z_{P}^\partial\hat{\mathcal O}_{P,R}$.

\end{document}